\documentclass[useAMS,usenatbib]{mn2e}

\usepackage{graphicx}
\usepackage{txfonts}
\usepackage{natbib}
\newcommand\cubi{${\it c}_{\rm U,B,I}$}
\newcommand\cubv{${\it c}_{\rm U,B,V}$}

\title[]{The SUMO project I. A survey of multiple populations in globular clusters
\thanks{Based on observations made with the INT telescope operated on 
the island of La Palma by the Isaac Newton Group in the Spanish Observatorio del 
Roque de los Muchachos of the Instituto de Astrof{\'i}sica de Canarias. Based on 
observations made with MPG/ESO 2.2m telescope in La Silla, Chile, proposal 088.A-9012.}}
\author[M.\, Monelli et al.]
{M.\, Monelli$^{1,2}$\thanks{E-mail:monelli@iac.es},
A.\, P.\, Milone$^{1,2,3}$,
P.\,B.\, Stetson$^{4}$,
A.\, F.\, Marino$^{3,5}$,
S.\, Cassisi$^{6}$,
\newauthor
A.\, Del Pino Molina$^{1,2}$,
M.\, Salaris$^{7}$,
A.\, Aparicio$^{1,2}$,
M.\, Asplund$^{3}$,
F.\, Grundahl$^{8}$,
\newauthor
G.\, Piotto$^{9,10}$,
A.\, Weiss$^{5}$,
R.\, Carrera$^{1,2}$,
M.\, Cebri\'an$^{1,2}$, 
S.\, Murabito$^{1,2}$,
\newauthor
A.\, Pietrinferni$^{6}$,
L.\, Sbordone$^{11}$ \\
$^{1}$Instituto de Astrof\'{i}sica de Canarias, Calle Via Lactea s/n, 38205 La Laguna, Tenerife, Spain \\
$^{2}$Departamento de Astrof\'{i}sica, Universidad de La Laguna, Tenerife, Spain \\
$^{3}$Research School of Astronomy \& Astrophysics, Australian National University, Mt Stromlo Observatory, via Cotter Rd, Weston, ACT 2611, Australia \\
$^{4}$Dominion Astrophysical Observatory, NRC-Herzberg, 5071 West Saanich Road, Victoria, BC V9E 2E7, Canada \\
$^{5}$Max-Planck-Institut f\"ur Astrophysik Karl-Schwarzschild-Str.\, 1 85741 Garching bei M\"unchen Germany  \\
$^{6}$INAF-Osservatorio Astronomico di Teramo, via M.\, Maggini, 64100 Teramo, Italy \\
$^{7}$Astrophysics Research Institute, Liverpool John Moores University, 12 Quays House, Birkenhead, CH41 1LD, UK\\
$^{8}$Department of Physics and Astronomy, Aarhus University, Ny Munkegade, 8000 Aarhus C, Denmark \\
$^{9}$Dipartimento di Fisica e Astronomia `Galileo Galilei', Universit\`a di Padova, Vicolo dell'Osservatorio 3, Padova, I-35122, Padova, Italy.\\
$^{10}$INAF-Osservatorio Astronomico di Padova, Vicolo dell'Osservatorio 5, Padova I-35122, Italy\\
$^{11}$Zentrum f\"{u}r Astronomie der Universit\"{a}t Heidelberg, Landessternwarte, K\"{o}nigstuhl 12, 69117 Heidelberg, Germany \\
 }

\begin{document}

%\date{Accepted xxx December 15. Received xxx December 14; in original form xx October 11}
\date{Revised Version Dec 06, 2012}

\pagerange{\pageref{firstpage}--\pageref{lastpage}} \pubyear{2012}

\maketitle

\label{firstpage}

\begin{abstract}  
We present a general overview and the first results of the
{\bf SUMO} project (a {\bf SU}rvey of {\bf M}ultiple p{\bf O}pulations in
Globular Clusters). The objective of  this survey is the study of multiple
stellar populations in the largest sample of  globular clusters
homogeneously analysed to date. To this aim we obtained high signal-to-noise
(S/N$>$50) photometry for main sequence stars with mass down to $\sim$0.5
$\mathcal{M}_{\rm odot}$ in a large sample of clusters using both archival
and proprietary $U$, $B$, $V$, and $I$ data from ground-based telescopes. \\
In this paper, we focus on the occurrence of multiple stellar populations in
twenty three clusters. We have defined a new photometric index
\cubi ~= $(U-B) - (B-I)$, that turns   out to be very effective for
identifying multiple sequences along the red giant branch (RGB). We found
that in the $V$-\cubi~ diagram all clusters  presented in this paper show
broadened or multimodal RGBs, with the presence of two or more  components.
We found a direct connection with the chemical properties of different
sequences, that display different abundances of light elements (O, Na, C, N,
and Al). The \cubi~ index is also a powerful tool to identify distinct 
sequences of stars along the horizontal branch and, for the first time in
the case of  NGC\,104 (47\,Tuc), along the asymptotic giant branch. Our
results demonstrate that  {\it i)} the presence of more than two stellar
populations is a common feature amongst globular clusters, as already
highlighted in previous work; {\it ii)} multiple sequences  with different
chemical contents can be easily identified by using standard Johnson
photometry obtained with ground-based facilities; {\it iii)} in the study of
GC multiple stellar populations the \cubi~ index is alternative to
spectroscopy, and has the advantage of larger statistics.
\end{abstract}

\begin{keywords}
globular clusters: general -- techniques: photometric
\end{keywords}

\section{Introduction}\label{sec:intro}

 There is an increasing body of evidence suggesting that both Galactic
and Magellanic Cloud globular clusters host multiple populations, a byproduct of 
the cluster internal chemical evolution. The majority of GCs appear to have 
at least two approximately coeval sub-populations, or rather two
generations of stars. The recognition of this fact came with the
increasing observational evidence for what was denoted as 
\lq{star-to-star abundance variations}\rq\ or \lq{abundance
anomalies}\rq, that manifest themselves as variations of nitrogen, carbon,
oxygen and sodium -- in some clusters also  magnesium and aluminum -- 
abundances amongst stars in individual clusters. 
These abundance differences are found in stars  of all
evolutionary stages \citep[see, e.g.\,][]{kraft92,cannon98,gratton01}.

This phenomenon has been known for a few decades
\citep[see, e.g.\,][]{osborn71,kraft78, cohen78,norris79}, but was  
limited to a small number of stars in each cluster. The striking, additional evidence
that GCs host multiple populations came from both high-quality photometry for
thousands of cluster stars, as well as from systematic spectroscopic investigation
thanks to modern multi-object spectroscopy.
As it turned out, multiple cluster sequences in the colour-magnitude diagrams 
(CMDs) show up in all the clusters studied with appropriate filters. 
The occurrence of multiple populations can be related to the evidence of 
main sequence (MS) splittings like in $\omega$ Centauri \citep{anderson97,bedin04},  
NGC\, 2808 \citep{dantona05,piotto07}, or NGC\, 6752 \citep{milone10}.
but it is also associated to the photometric detection of multimodal RGBs  
\citep[e.g.\,][]{grundahl99,marino08,yong08,lee09},
and in some cases of a sub-giant branch (SGB) splittings 
\citep[e.g.\,][]{milone08,marino09,piotto12}.

The most recent photometric analyses have shown that the CMDs of GCs 
consist of intertwined sequences of the two or more populations, 
whose separate identities can be followed continuously from the MS up to the red giant 
branch (RGB), and thence to the horizontal branch \citep[HB, NGC\,104, ][]{milone12d}.

Photometric analyses have clearly shown that the use of photometric filters
sampling the ultra-violet (UV) portion of the stellar spectra, is a very
powerful tool for detecting and tracing multiple populations in the CMDs of 
GCs \citep[e.g.\,][]{grundahl99,marino08}.

Computations of suitable model atmospheres \citep{sbordone11} accounting for
the peculiar chemical patterns of second generation stars have explained  
the pivotal role played by UV photometric filters in  studying
sub-populations in GCs. The UV portion of  stellar spectra is hugely
affected by the change in the abundance of CN and NH molecules, due to the
light-element anticorrelations observed in the stars belonging  to the
second stellar generation; using suitable combinations of photometric
filters, including UV filters, allows one to emphasize the difference in the
radiative flux between primordial cluster stars  and those enriched by the
high-temperature H-burning products.

Obviously, (high-resolution) spectroscopy allows a more detailed analysis of
the  specific chemical patterns associated to  second generation stars; but
for the majority of GCs this kind of analysis is limited to the brightest
RGB stars, and in any case to a limited number of objects.

Tracing the multiple population phenomenon in a large fraction of cluster
stars (if not in the whole cluster)  is mandatory to establish the actual
fraction of second generation stars in individual GCs.  In addition, the
comparison between photometry  and theoretical stellar evolutionary models
\citep[e.\ g.][]{pietrinferni09} and colour - $T_{\rm eff}$ transformations,
that   account for the specific chemical patterns of the distinct
sub-populations  \citep[see ][]{sbordone11}, allows to  investigate how
stellar structural and evolutionary properties  are affected by the chemical
abundance distribution  of each distinct sub-population.

In this work we present an overview of the SUMO project,  -- a SUrvey of
Multiple pOpulations --, together with its first results. The paper is
organized as follows: \S \ref{sec:project} introduces the main objectives of
the survey, while the data and the  derived CMDs are presented in \S
\ref{sec:data} and \S \ref{sec:selections}. \S \ref{sec:uband} introduces a
new photometric index, based on the combination of  three bands, that is
effective to highlight different sequences along the RGB.  The connection
with the chemical content of sub-populations identified in this way
is discussed in \S \ref{sec:spec}. Our conclusions in \S
\ref{sec:conclusions} close the paper.

\section{The SUMO project}\label{sec:project}

The basic idea of the SUMO project %-- a SUrvey of Multiple pOpulations -- 
is to create a database of homogeneous, wide-field, multiwavelength, photometric data
of GCs. The main aim is to study the multiple population phenomenon, and in particular
the primary objectives of the survey are:\\
$\bullet$ investigate the occurrence of multiple populations in a large sample of GCs.
Is this a property common to {\itshape all} GCs?\\
$\bullet$ Identify multiple populations along different evolutionary phases, from the MS to
the RGB and HB;\\
$\bullet$ verify the connection between photometric and spectroscopic properties of
different stellar populations;\\
$\bullet$ study the radial gradients over a large fraction of the body of each cluster.

$U$,$B$,$V$,$I$ photometry for a large number of clusters will also be used to study a number of
different topics, to be presented in forthcoming papers: \\
{\itshape i)} the effect of differential reddening, if present, following the 
procedure described in \citet[][see also Sect. \ref{sec:reddening}]{milone09}; \\
{\itshape ii)} the population of blue stragglers, and their radial gradients; \\
{\itshape iii)} the binary fraction, which will complement similar works on the same 
GCs based on Hubble Space Telescope ({\it HST}) data and limited to the cluster 
centres \citep[e.\ g.][]{sollima07,milone08}. \\
{\itshape iv)} the properties of the RGB bump over a large metallicity range; \\
{\itshape v)} a self-consistent study of the HB morphology; \\
{\itshape vi)} identification of spectroscopic targets for future follow-ups; \\
{\itshape vii)} new theoretical models will have to be developed, with ad-hoc chemical mixtures
and the proper model atmospheres, following the pioneering work of \citet{sbordone11}.

%%%%%%%%%%%%%%%%%%%%%%%%%%%%%%%%%%%%%% FIG 1 %%%%%%%%%%%%%%%%%%%%%%%%%%%%%%%%%%%
\begin{centering}
\begin{figure*}
 \includegraphics[width=12cm]{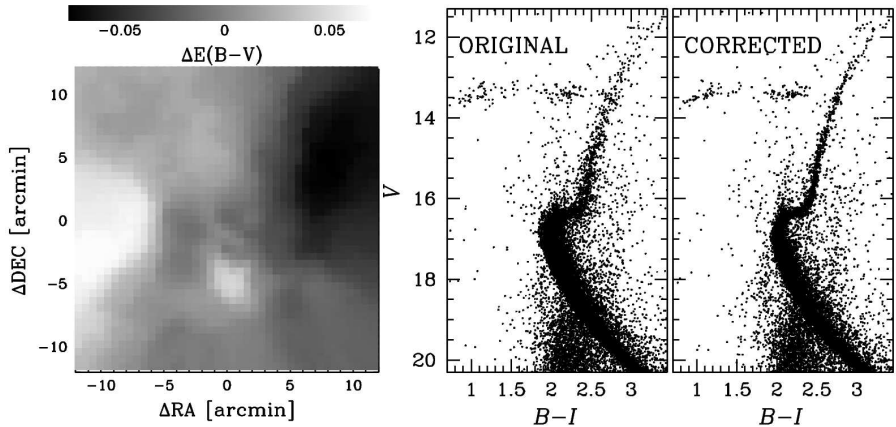}
 \caption{{\textit Left-}Reddening map for NGC\,6121. {\textit Center and Right-}
 $V$ {\em vs} $B-I$ CMD before and after applying the correction for differential reddening.}
 \label{fig:redde}
\end{figure*}
\end{centering}
%%%%%%%%%%%%%%%%%%%%%%%%%%%%%%%%%%%%%%%%%%%%%%%%%%%%%%%%%%%%%%%%%%%%%%%%%%%%%%%%

This survey has been designed to complement the {\it HST} data available in
the literature,  that have been successfully used to disentangle multiple
sequences in the CMD of the inner, most  crowded regions of a large number
of GCs \citep{piotto07,milone08,bellini10,milone12d,piotto12}. The large
field-of-view of ground based  data is mandatory to extend this result over
the main body of each cluster, tracing the  presence and the relative
contribution of distinct components. This is crucial to shed light on the 
formation mechanism of the multiple populations. In fact, theoretical models
\citep{dercole08} predict that a large fraction (up to 90\% or more) of
first-generation stars should be lost early on during the GC evolution,
because of the expansion and stripping of the cluster outer layers resulting
from the mass loss consequent to massive SNe explosions. The fact that
first-generation stars were much more numerous at the time of the cluster
formation can also account for the polluting material needed to explain the
following stellar generation.  Signatures of these dynamical evolution are
expected to be observable in the present-day  stellar radial distribution,
at least in clusters with long relaxation times; a systematic  study of the
radial gradients would therefore provide solid constraints to understand the
nature of the polluters and  the processes that have caused the formation of
successive stellar generations. 

Finally, following previous works \citep{marino08, milone12d}, we emphasize
the use  of the $U$-band as an efficient diagnostic to identify multiple
stellar populations.  For this reason, from the observational point of view
(see \S \ref{sec:data}), we devoted a significant fraction of our observing
runs collecting data in this filter. The colour combination presented in
this paper (see \S. \ref{sec:uband})  turned out to be very effective to
show the presence of multiple populations along the RGB, correlating   with
the chemical properties of stars. In the following section, we summarize the
properties of the target clusters.

	\subsection{Public products}\label{sec:products}

 One of the objectives of this project is to disseminate
the results to the greatest extent possible, releasing to the general community a large number of
products. With this aim, we set a web page, {\itshape http://www.iac.es/project/sumo} 
which will be regularly updated with: plots
(CMDs, colour-colour), reddening maps, fiducial lines.

\begin{table*}
  \caption{Summary of data for the 23 clusters included in this papers. We summarize the cluster name, coordinate,
  and the maximum number of images, in each band, a given stars has been measured in.\label{tab:tab01}}
  \begin{tabular}{l|c|c|c|r|c|r|c|r|c|r|}
  \hline 
 Cluster             &  R.A.            & Dec.             & U   &  B    &  V    &  I	\\
                     &                  &                  &     &       &       & 	 \\
 \hline
NGC\,104 [47 Tuc]    &  00 24 05.67     &    -72 04 52.6   & 21  &  106  &  115  &  103 \\
NGC\,288	     &  00 52 45.24     &    -26 34 57.4   &  9  &   63  &  100  &   68	 \\
NGC\,362             &  01 03 14.26     &    -70 50 55.6   & 11  &  140  &  162  &  151	 \\
NGC\,2808            &  09 12 03.10     &    -64 51 48.6   & 48  &  652  &  545  &  203  \\
NGC\,3201	     &  10 17 36.82     &    -46 24 44.9   & 13  &    4  &    4  &    4	 \\
NGC\,4590  [M~68]    &  12 39 27.98     &    -26 44 38.6   & 14  &   48  &   48  &   35	 \\
NGC\,5904  [M~5]     &  15 18 33.22     &    +02 04 51.7   & 28  &   75  &  132  &  127  \\
NGC\,6093  [M~80]    &  16 17 02.41     &    -22 58 33.9   & 21  &   25  &   45  &   22	 \\
NGC\,6121  [M~4]     &  16 23 35.22     &    -26 31 32.7   & 12  & 1026  & 1425  &   41	 \\
NGC\,6205  [M~13]    &  16 41 41.24     &    +36 27 35.5   & 20  &   58  &   54  &   67  \\
NGC\,6218  [M~12]    &  16 47 14.18     &    -01 56 54.7   & 46  &  196  &  212  &  166  \\
NGC\,6254  [M~10]    &  16 57 09.05     &    -04 06 01.1   & 17  &   18  &   27  &   29  \\
NGC\,6366	     &  17 27 44.24     &    -05 04 47.5   &  8  &    9  &   30  &   18  \\
NGC\,6397            &  17 40 42.09     &    -53 40 27.6   & 11  &   42  &   36  &   28	 \\
NGC\,6541            &  18 08 02.36     &    -43 42 53.6   & 12  &   33  &   36  &   23  \\
NGC\,6681  [M~70]    &  18 43 12.76     &    -32 17 31.6   & 13  &   28  &   48  &   38	 \\
NGC\,6712            &  18 53 04.30     &    -08 42 22.0   & 35  &   38  &   49  &  ---  \\
NGC\,6752            &  19 10 52.11     &    -59 59 04.4   & 35  &   84  & 1176  &   28	 \\
NGC\,6809  [M~55]    &  19 39 59.71     &    -30 57 53.1   & 12  &   40  &   40  &   36  \\
NGC\,6934	     &  20 34 11.37     &    +07 24 16.1   & 15  &   38  &   42  &   39	 \\
NGC\,6981  [M~72]    &  20 53 27.70     &    -12 32 14.3   &  6  &  241  &  277  &  218	 \\
NGC\,7078  [M~15]    &  21 29 58.33     &    +12 10 01.2   & 31  &  277  &  271  &  196	 \\
NGC\,7099  [M~30]    &  21 40 22.12     &    -23 10 47.5   &  9  &   38  &   48  &   20  \\
\hline
\end{tabular}
\end{table*}

%%%%%%%%%%%%%%%%%%%%%%%%%%%%%%%%%%%%%% FIG 2 %%%%%%%%%%%%%%%%%%%%%%%%%%%%%%%%%%%
\begin{centering}
\begin{figure*}
 \includegraphics[]{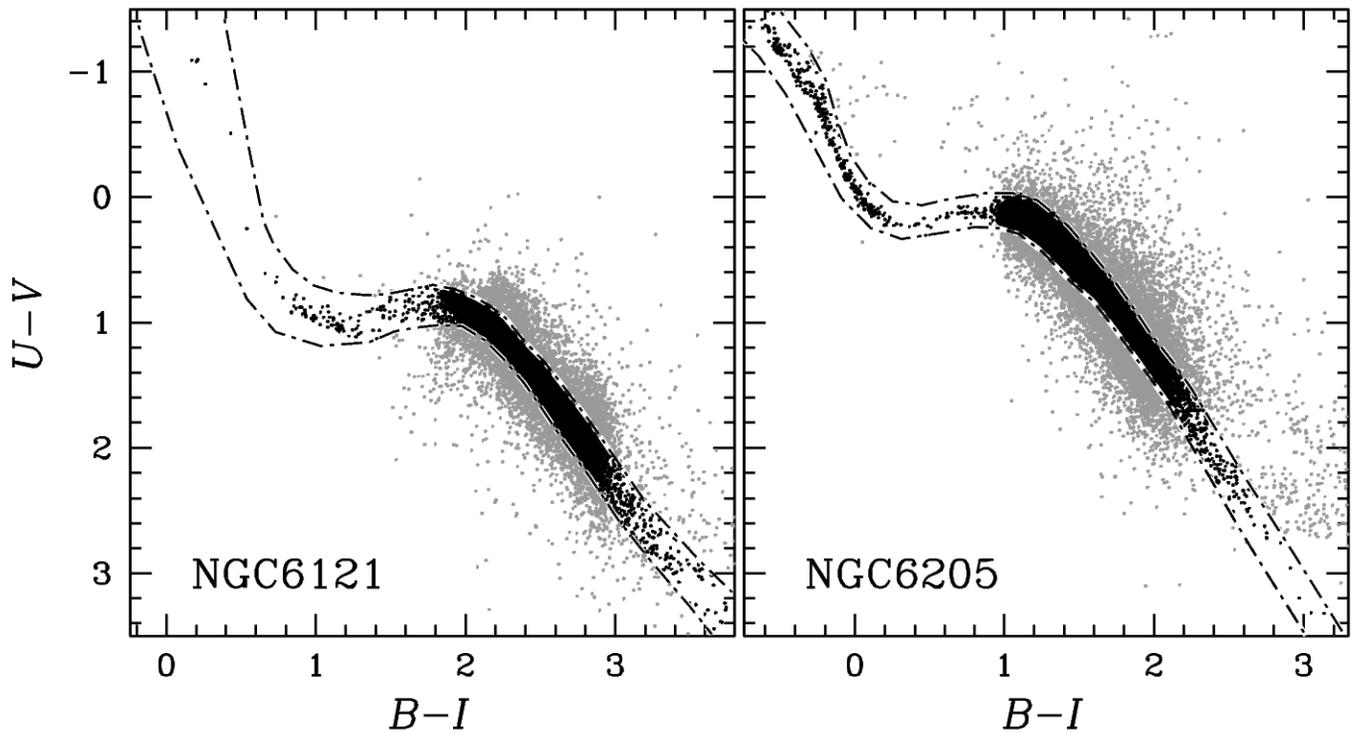}
 \caption{($U-V$) {\em vs} ($B-I$) colour-colour plane for the NGC\,6121 (left) and NGC\,6205 
 fields (right). Following \citet{bono10} we use this diagram to identify foreground Galactic
 stars and background unresolved galaxies. The bona-fide cluster stars are selected inside the
 area marked by the overplotted line.}
 \label{fig:colcol}
\end{figure*}
\end{centering}
%%%%%%%%%%%%%%%%%%%%%%%%%%%%%%%%%%%%%%%%%%%%%%%%%%%%%%%%%%%%%%%%%%%%%%%%%%%%%%%%

%%%%%%%%%%%%%%%%%%%%%%%%%%%%%%%%%%%%%% FIG 3 %%%%%%%%%%%%%%%%%%%%%%%%%%%%%%%%%%%
\begin{figure*}
 \includegraphics[]{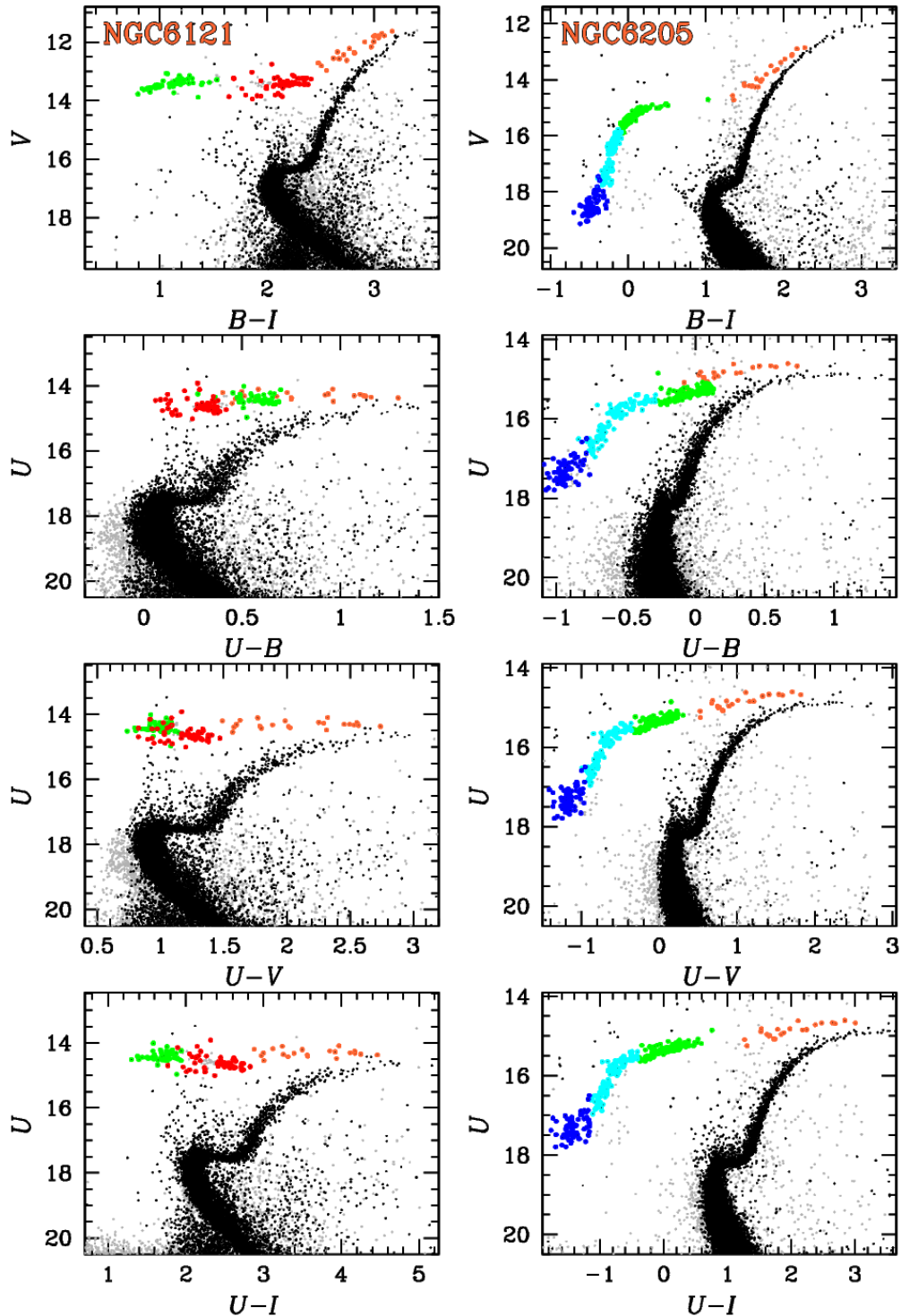}
 \caption{Selection of different CMDs of NGC\,6121 and NGC\,6205.
 The plot shows that our photometry reaches well below the MS turn off, with good 
 accuracy in all four bands. AGB and HB stars are highlighted with different
 colours to show their behaviour in the CMDs.}
 \label{fig:cmds}
\end{figure*}
%%%%%%%%%%%%%%%%%%%%%%%%%%%%%%%%%%%%%%%%%%%%%%%%%%%%%%%%%%%%%%%%%%%%%%%%%%%%%%%%

%%%%%%%%%%%%%%%%%%%%%%%%%%%%%%%%%%%%%% FIG 4 %%%%%%%%%%%%%%%%%%%%%%%%%%%%%%%%%%%
\begin{centering}
\begin{figure*}
 \includegraphics[width=12cm]{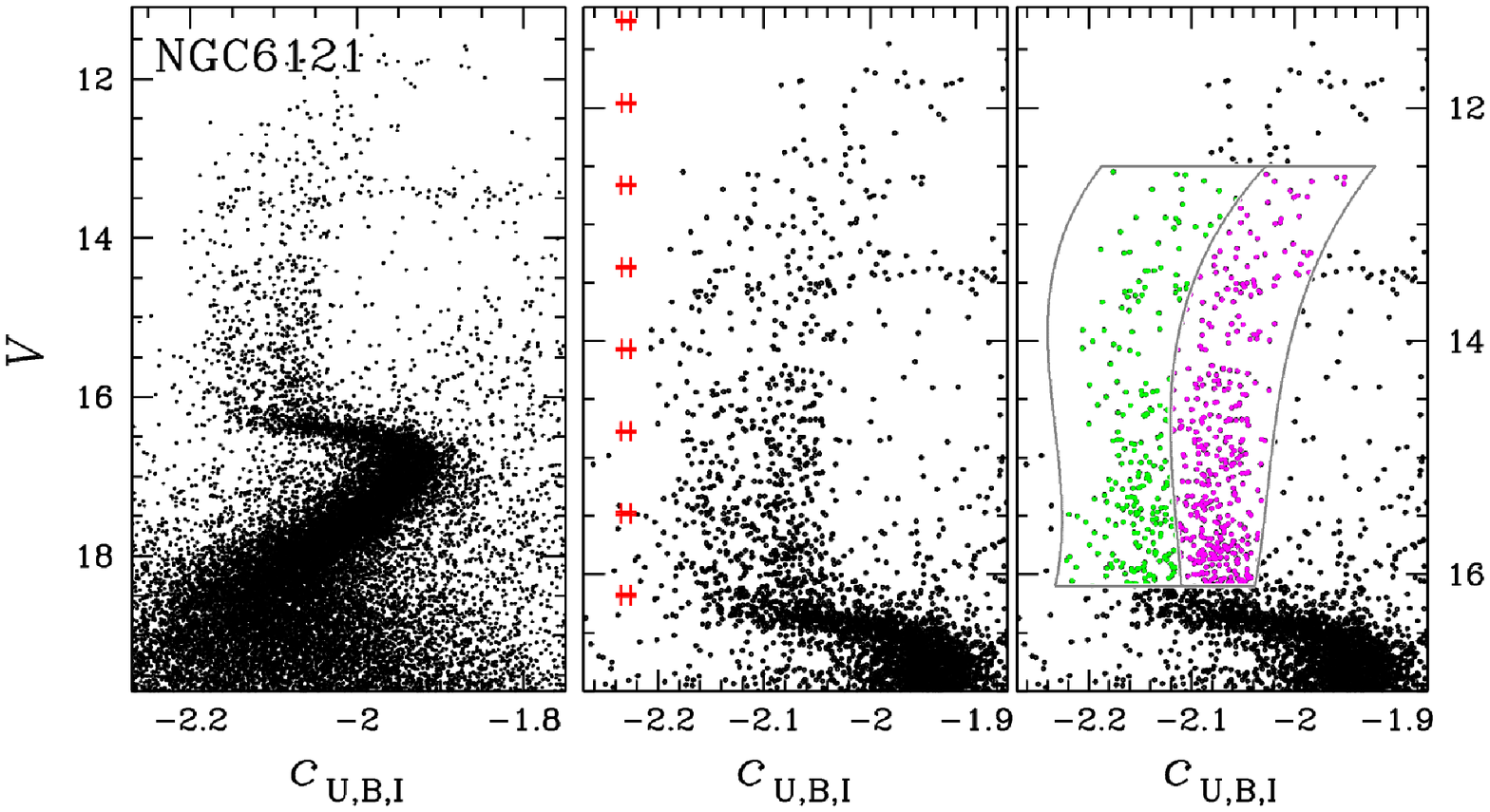}
 \includegraphics[width=12cm]{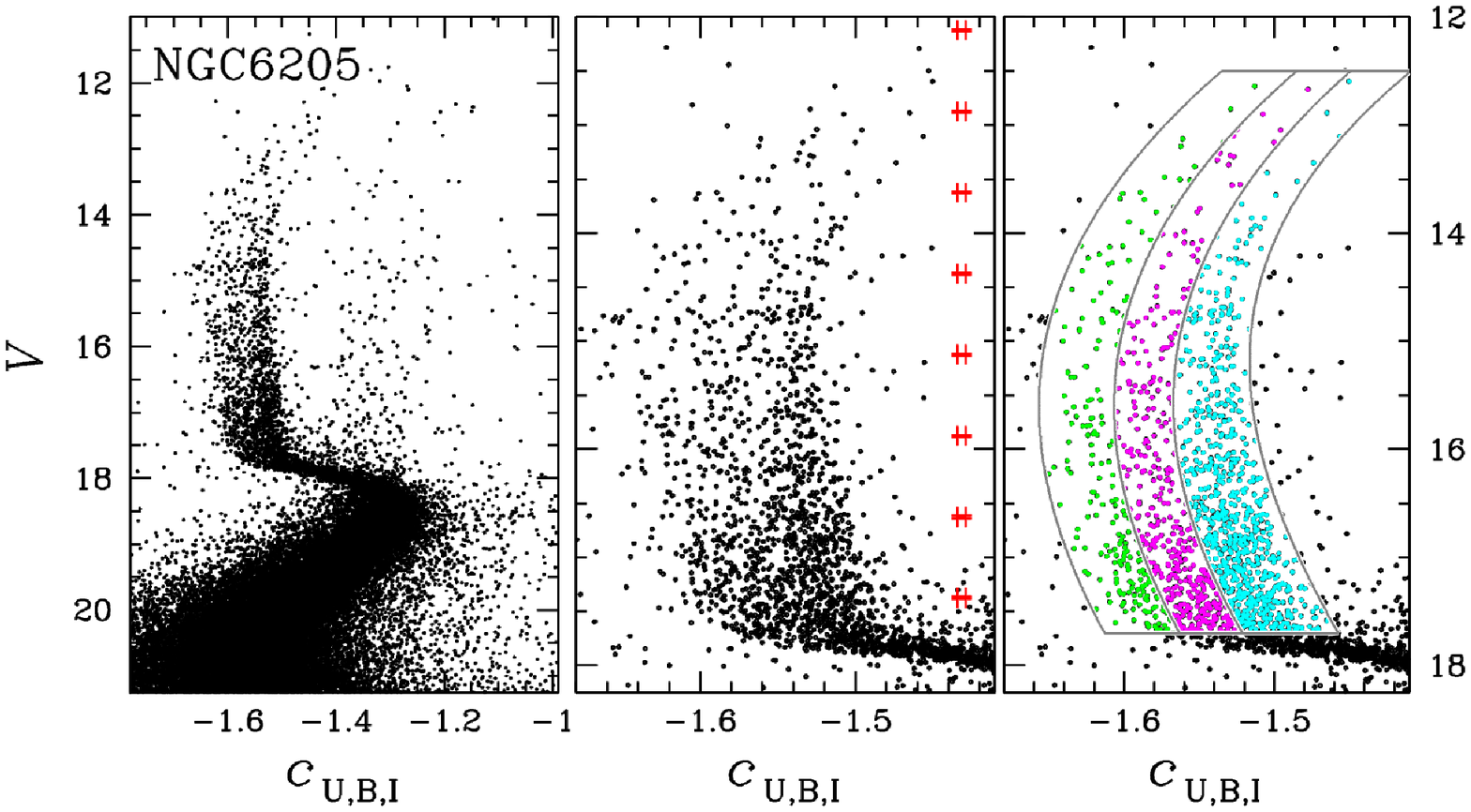}
 \caption{NGC\,6121 and NGC\,6205 are adopted to illustrate the properties
 of the ($V$, \cubi) pseudo-CMDs. The left and central panels show the entire diagram
 and a zoom in the RGB region, respectively. Using this diagram, we identified two and
 three subpopulations, respectively, highlighted in the right panels.
  }
 \label{fig:cubi6205}
\end{figure*}
\end{centering}
%%%%%%%%%%%%%%%%%%%%%%%%%%%%%%%%%%%%%%%%%%%%%%%%%%%%%%%%%%%%%%%%%%%%%%%%%%%%%%%%

	\subsection{Properties of the target clusters}\label{sec:targets}

\begin{centering}
\begin{figure*}
 \includegraphics[width=16cm]{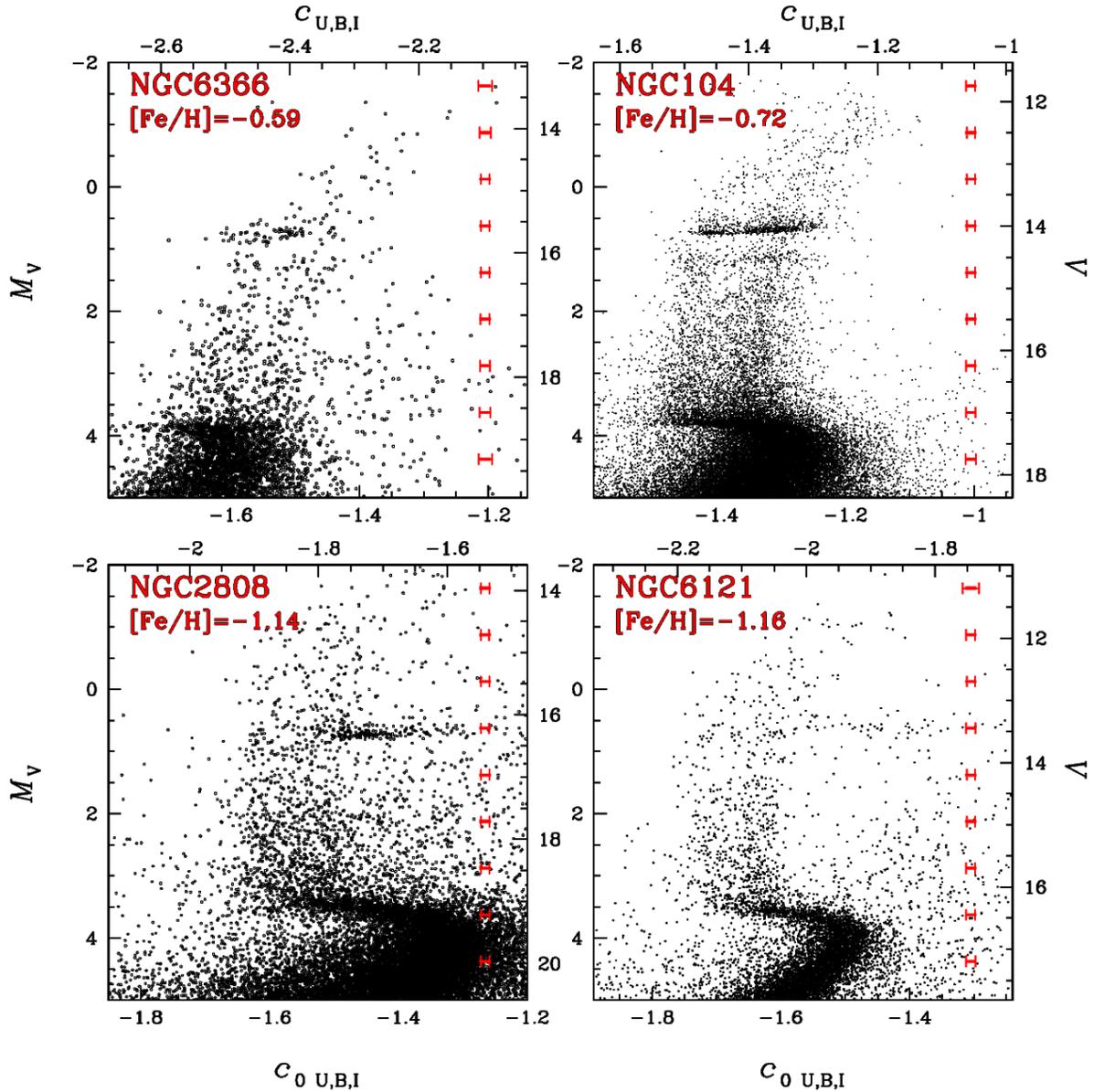}
 \caption{
 $M_V$ {\em vs} \cubi~ CMDs for NGC\,6366, NGC\,104, NGC\,2808, and NGC\,6121. 
 The split of the RGB is evident all the clusters. We adopted the distance and mean 
 reddening  from \citet{harris10}  to shift the diagrams to absolute magnitudes.
 Apparent magnitude and colour are labeled in the right and top axes, respectively.
 The clusters are sorted in order of increasing metallicity.}
 \label{fig:cubi}
\end{figure*}
\end{centering}

\begin{centering}
\begin{figure*}
 \includegraphics[width=16cm]{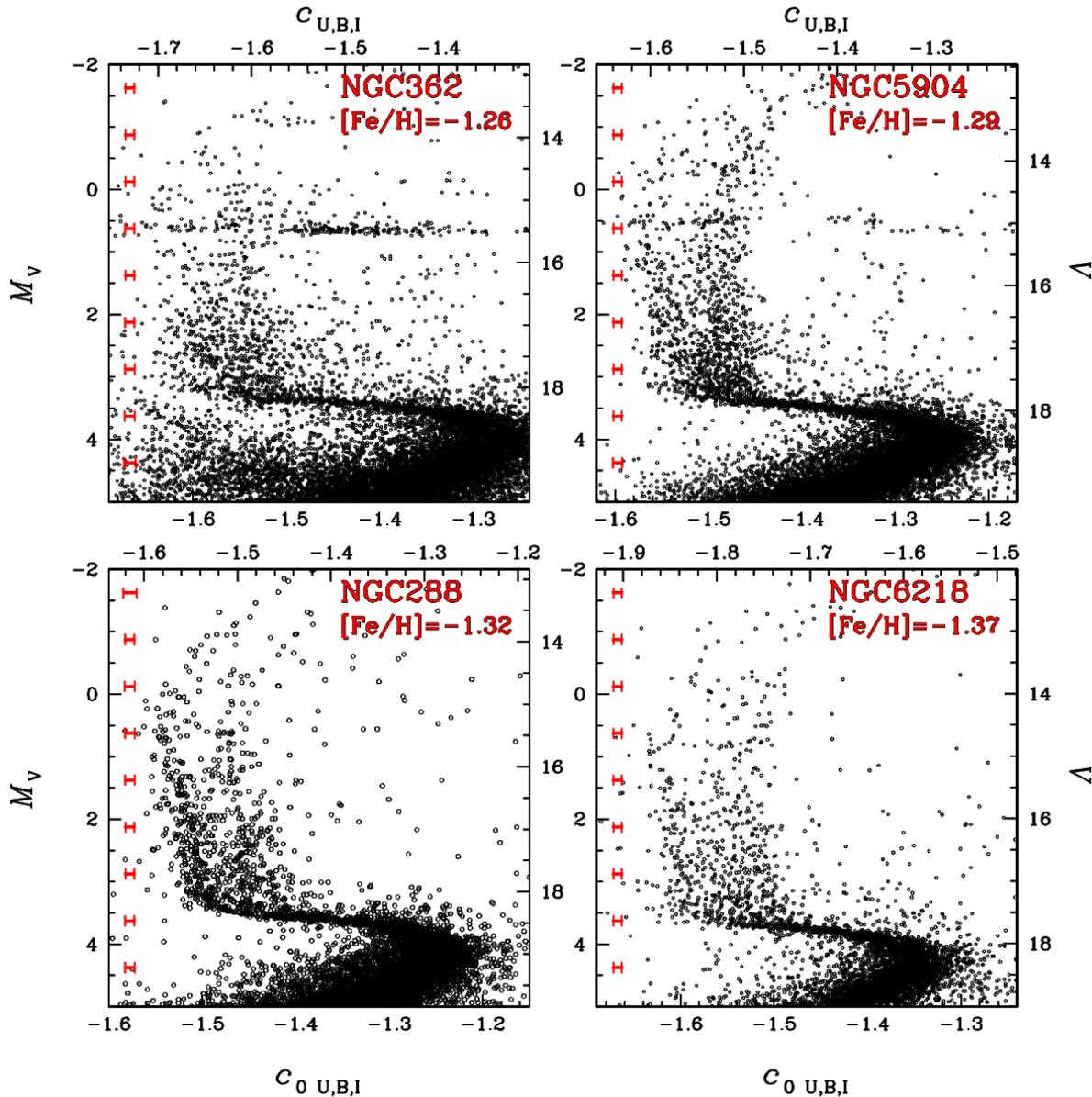}
 \caption{Same as Fig.\ref{fig:cubi}, but for NGC\,362, NGC\,5904, NGC\,288, and NGC\,6218.}
 \label{fig:cubi2}
\end{figure*}
\end{centering}

\begin{centering}
\begin{figure*}
 \includegraphics[width=16cm]{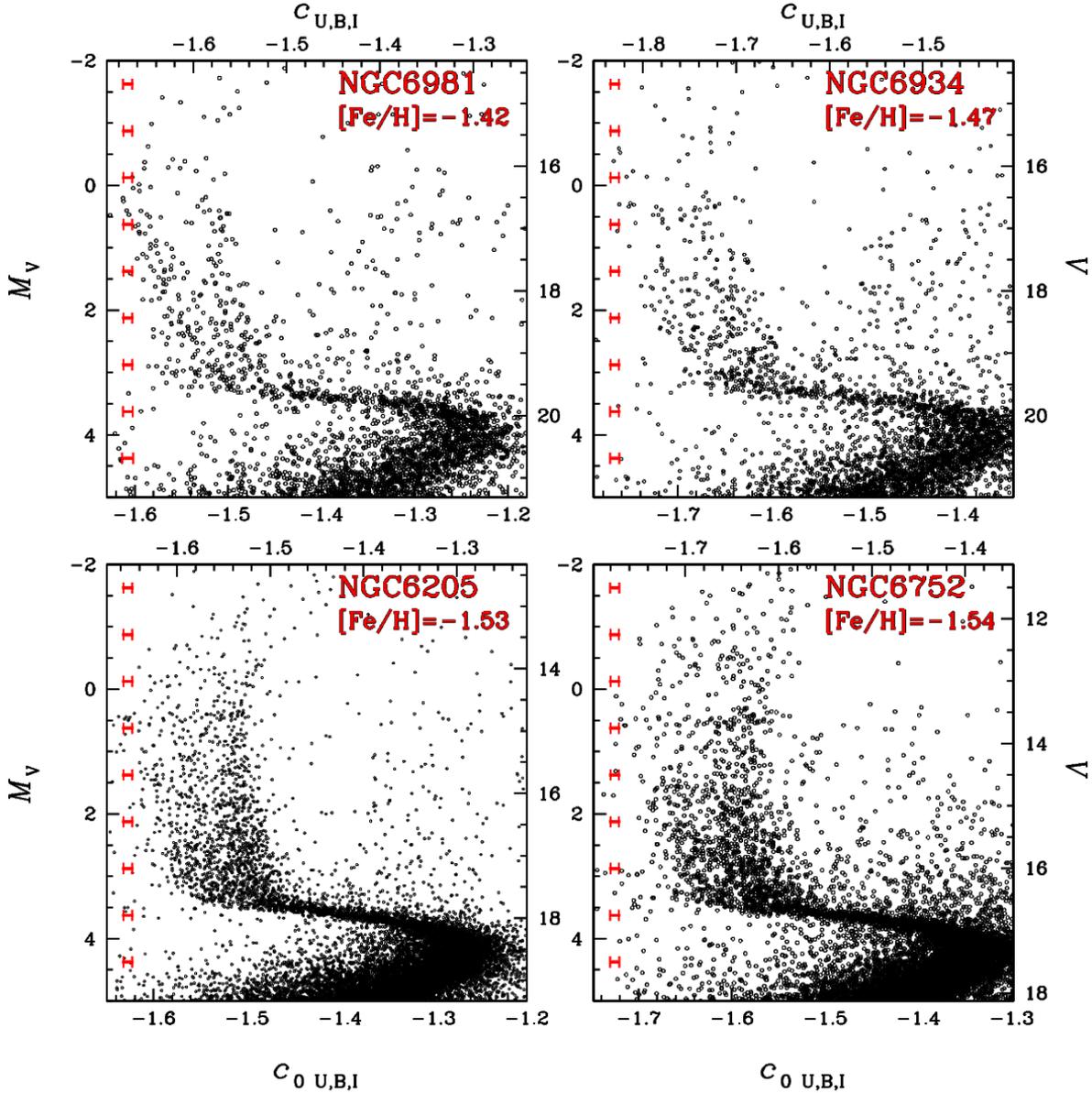}
 \caption{Same as Fig.\ref{fig:cubi}, but for NGC\,6981, NGC\,6934, NGC\,6205, and NGC\,6752.}
 \label{fig:cubi3}
\end{figure*}
\end{centering}

\begin{centering}
\begin{figure*}
 \includegraphics[width=16cm]{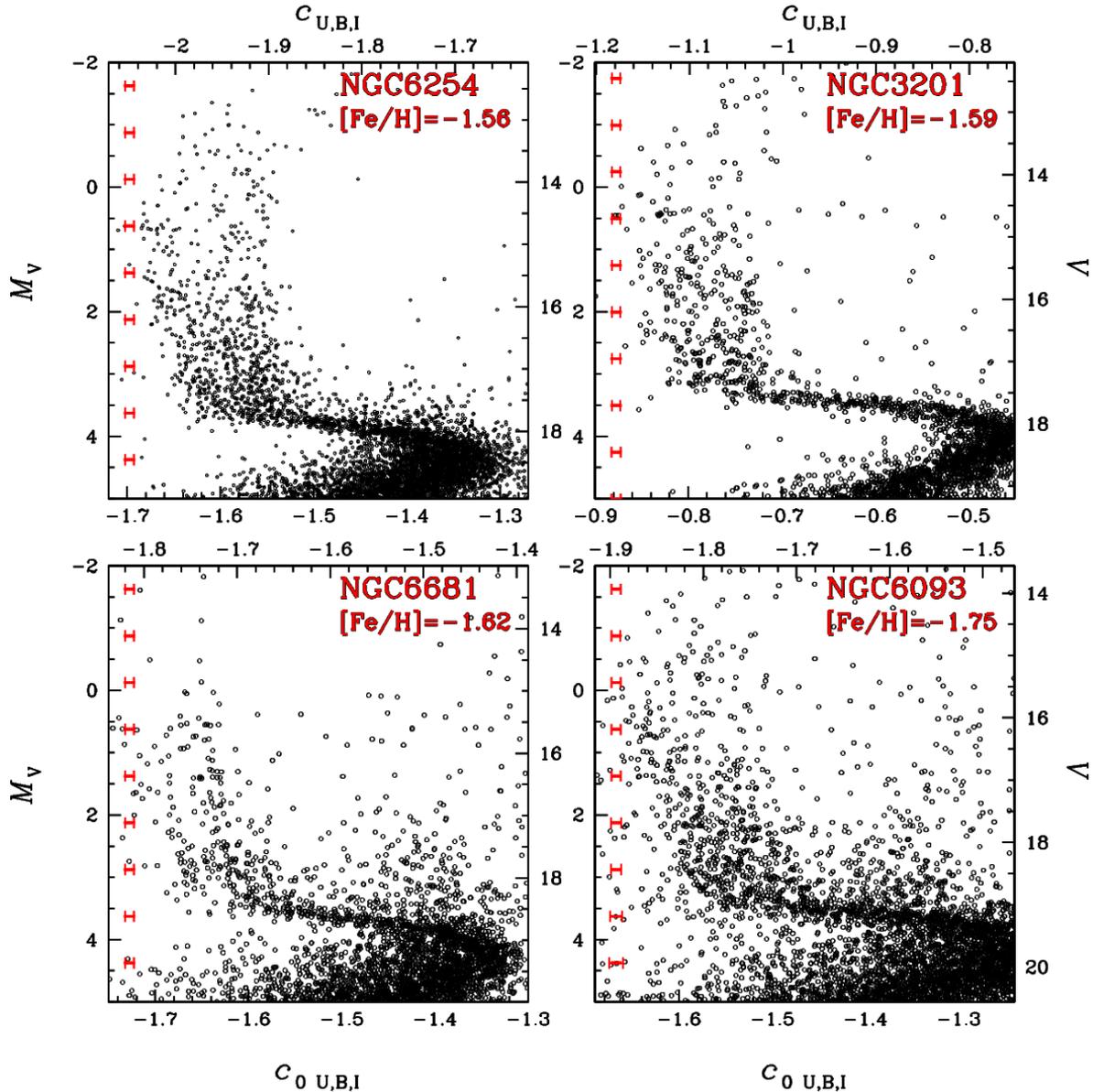}
 \caption{Same as Fig.\ref{fig:cubi}, but for NGC\,6254, NGC\,3201, NGC\,6681, and NGC\,6093.}
 \label{fig:cubi4}
\end{figure*}
\end{centering}

\begin{centering}
\begin{figure*}
 \includegraphics[width=16cm]{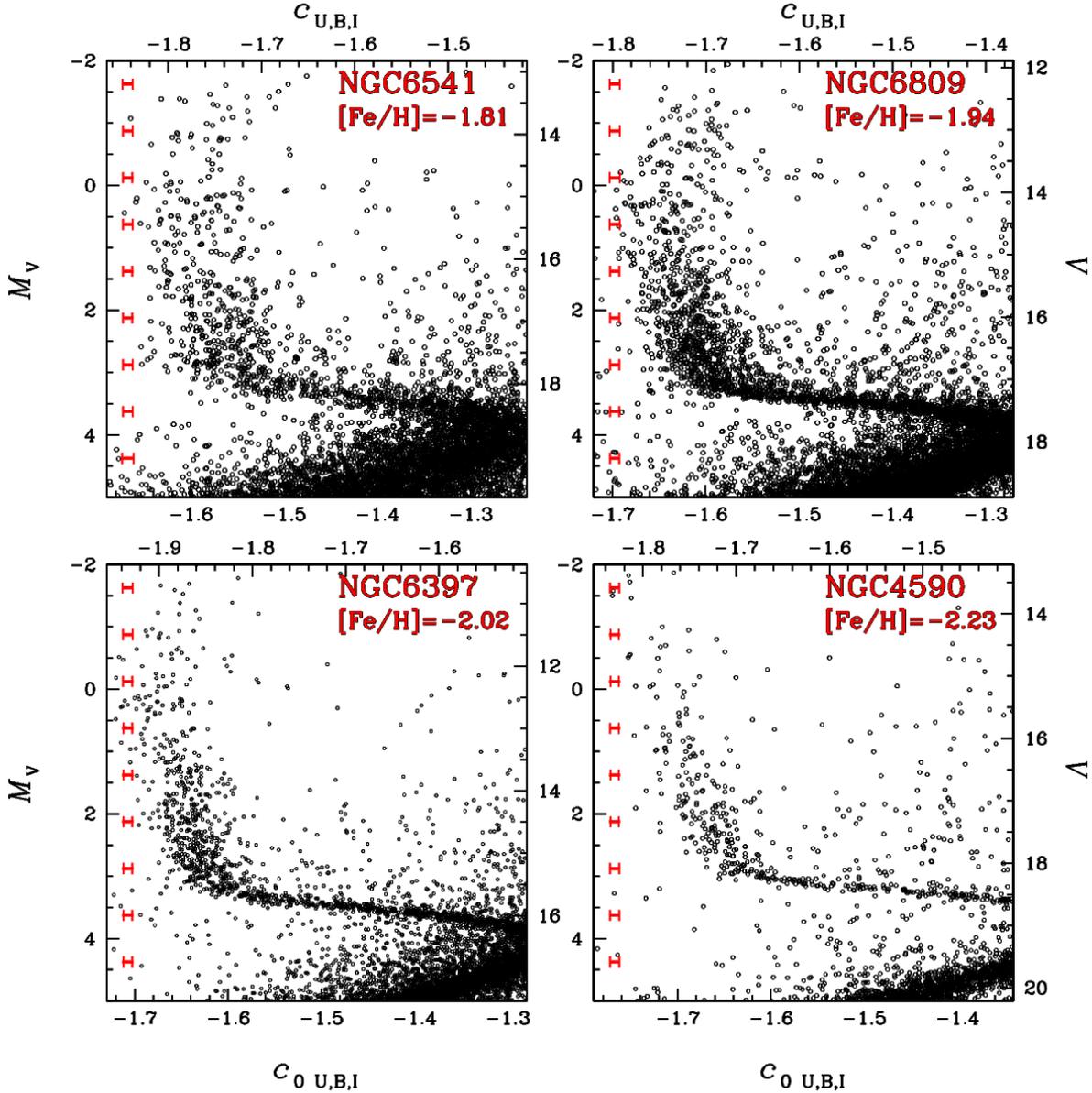}
 \caption{Same as Fig.\ref{fig:cubi}, but for NGC\,6541, NGC\,6809, NGC\,6397, and NGC\,4590.}
 \label{fig:cubi5}
\end{figure*}
\end{centering}

\begin{centering}
\begin{figure*}
 \includegraphics[width=16cm]{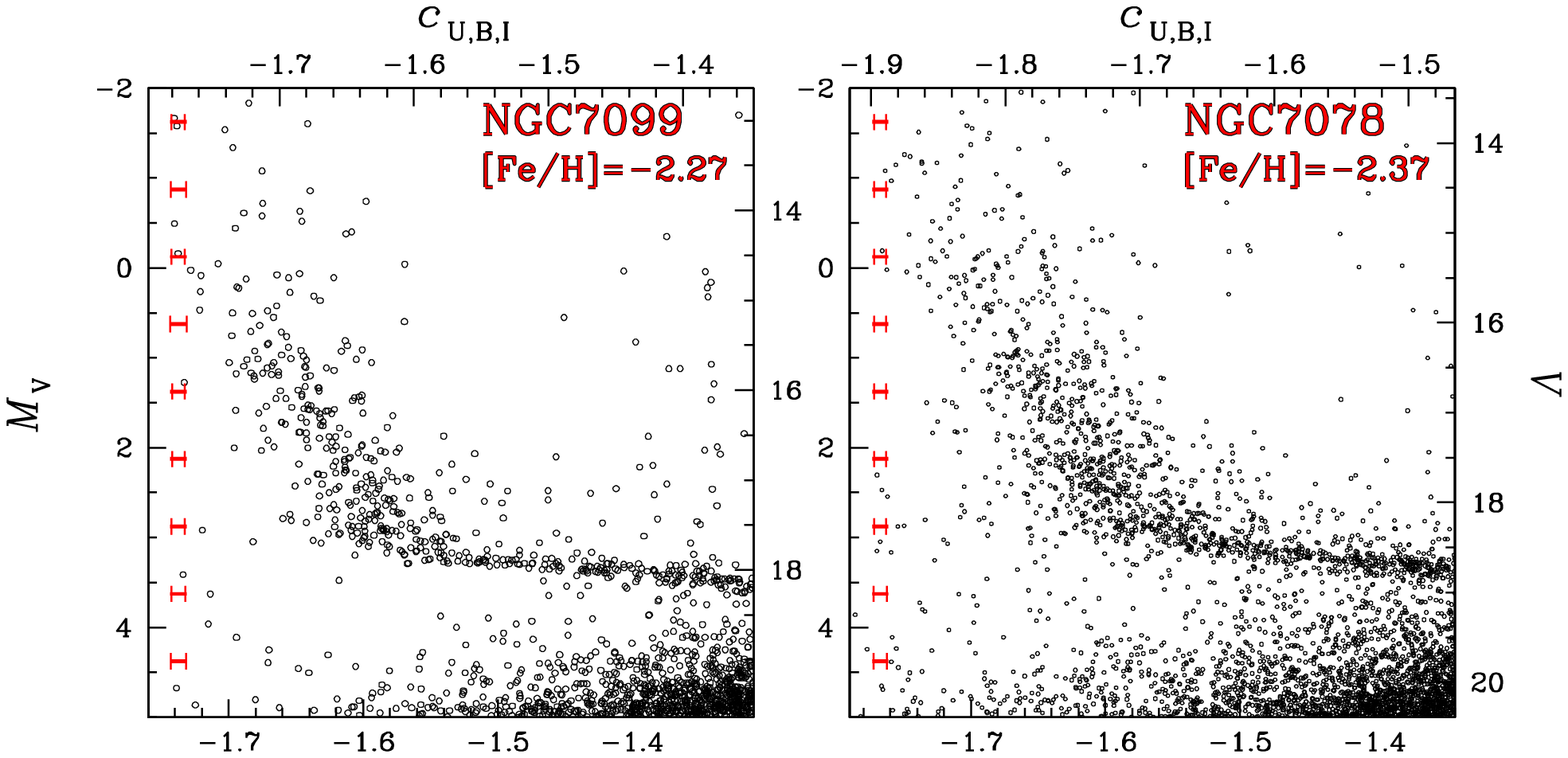}
 \caption{Same as Fig.\ref{fig:cubi} but for NGC\,7099 and NGC\,7078.}
 \label{fig:cubi6}
\end{figure*}
\end{centering}

The cluster sample presented in this paper includes both objects where
different generations of stars have been already discovered and widely
studied, and objects with no evidence of multiple stellar populations.  In
this Section we summarize the observational scenario. The main properties
will be used in the following to interpret the multimodal RGBs of
Fig.~\ref{fig:cubi}-\ref{fig:cubi6}.

The CMD of {\bf 47 Tuc-NGC\,104} consists of intertwined sequences, whose
separate identities can be followed continuously from the MS up to the RGB,
and thence to the HB \citep{milone12d}. The observed colours are
consistent with a pair of populations with different content of light
elements \citep[e.g.][]{norris79,cannon98,harbeck03,pancino10},  and small
helium differences \citep[e.g.][]{dicriscienzo11, nataf11, milone12e}. A
third population is visible only along the SGB, and includes less than 10\%
of the stars  \citep{anderson09, piotto12}.

%%%%%%%%%%%%%%%%%%%%%%%%%%%%%%%%%%%%%% FIG 5 %%%%%%%%%%%%%%%%%%%%%%%%%%%%%%%%%%%
\begin{centering}
\begin{figure*}
 \includegraphics[width=16cm]{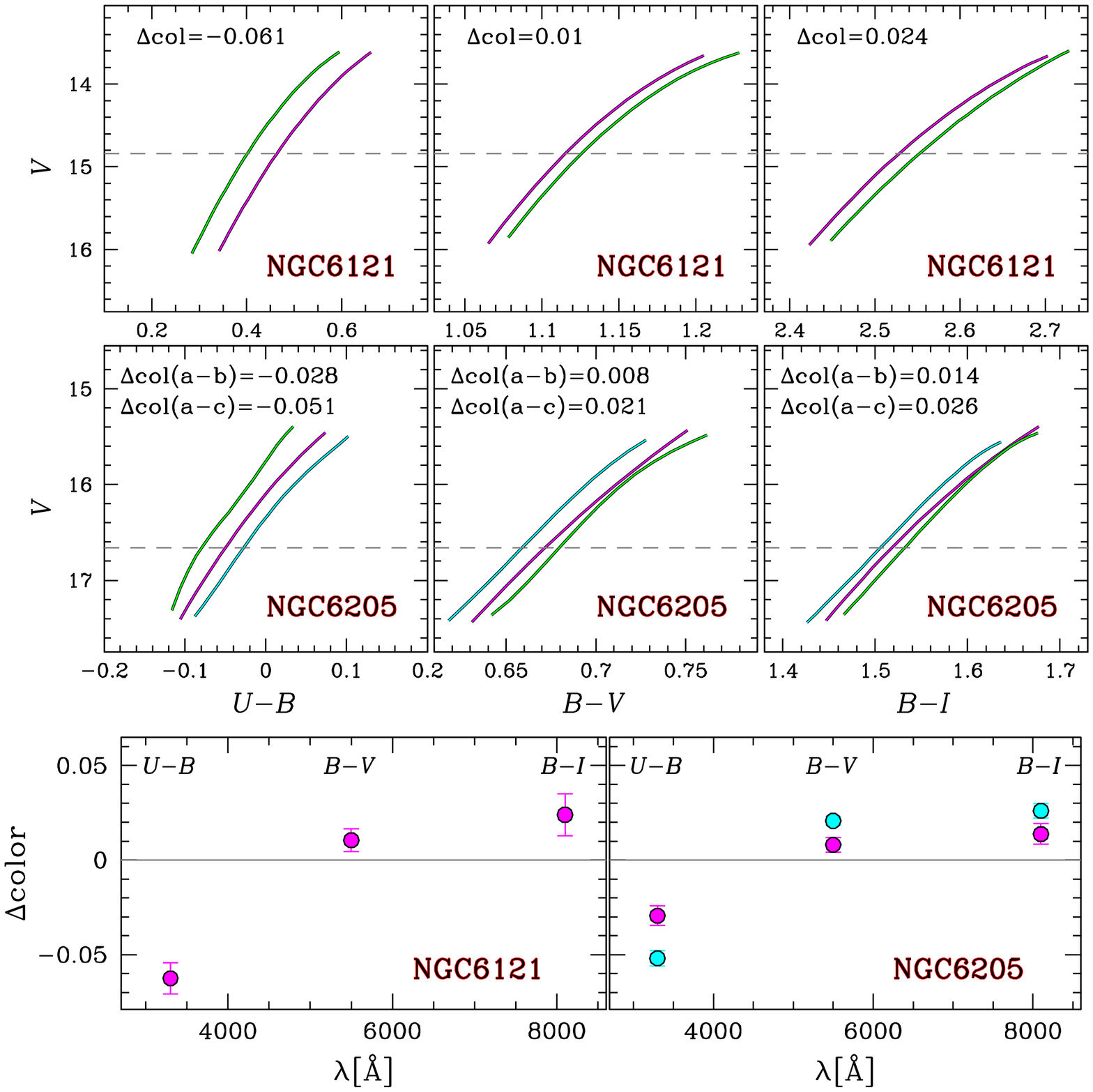}
 \caption{The ridge lines of the three populations detected along the RGB of NGC\,6205
 are used to show the efficiency of the \cubi~ index in identify multiple sequences. 
 The top three panels show the ridge lines in three different CMDs. 
 The bottom panels show the colour difference with respect of the green line, 
 calculated at $V = 14.82$ mag  and $V$=16.63 mag for the two clusters. The ($U-B$)-
 ($B-I$) index maximizes the colour difference, thus best 
 disentangling the different sequences.
 }
 \label{fig:works}
\end{figure*} 
\end{centering}
%%%%%%%%%%%%%%%%%%%%%%%%%%%%%%%%%%%%%%%%%%%%%%%%%%%%%%%%%%%%%%%%%%%%%%%%%%%%%%%%

%%%%%%%%%%%%%%%%%%%%%%%%%%%%%%%%%%%%%% FIG 6 %%%%%%%%%%%%%%%%%%%%%%%%%%%%%%%%%%%
%%%%%%%%%%%%%%%%%%%%%%%%%%%%%%%%%%%%%%%%%%%%%%%%%%%%%%%%%%%%%%%%%%%%%%%%%%%%%%%

{\bf NGC\,6121} is known to be a metal-intermediate GC, with
[Fe/H]$\approx-$1.1 and  [$\alpha$/Fe]$\approx$0.4.  Spectroscopy of bright
RGB stars revealed a significant spread in light-element abundances
\citep{gratton86, brown90, brown92, drake92, smith05, villanova11} with a
bimodal CN distribution \citep{norris81}, an O-Na  anticorrelation and a
Na-Al correlation \citep{ivans99, marino08}. The analysis of high-resolution
VLT/UVES spectra for about one hundred RGB stars revealed a bimodality in
the stellar distribution in the Na versus O plane \citep{marino08}.  The two
groups of Na-rich/O-poor, and Na-poor/O-rich stars populate two different
sequences along the RGB in the $U$ versus $U-B$ CMD. Na-rich stars define a
sequence on the red side of the RGB, while Na-poor stars populate a bluer,
more spread sequence.  Interestingly, NGC\,6121 hosts also a bimodal HB,
which is well populated both on the red and the blue side of the RR-Lyrae
gap. HB stars exhibit a bimodal Na and O distribution similar to what found
for RGB stars, with red-HB stars having solar Na and blue-HB stars being
oxygen depleted and sodium enhanced (\citealt{marino11},  see also
\citealt{norris81, smith93}). 

Among clusters with multiple stellar populations, {\bf NGC\,2808} is
certainly one of the most intriguing objects. Its CMD shows three distinct
MSs \citep{piotto07}, with the middle and the  blue MS being highly enhanced
in helium compared to the red MS -- $\Delta$Y$\sim$0.07 and
$\Delta$Y$\sim$0.13, respectively \citep{dantona05, piotto07, milone12e}. 
It also hosts a multimodal HB greatly extended blue-ward \citep{sosin97,
bedin00},  and a spread RGB \citet{lee09}. Furthermore, spectroscopic
studies of RGB, HB, and bright MS stars have revealed significant
star-to-star differences in the light-element abundances and an extended Na-O
anticorrelation \citep{norris81, carretta06, bragaglia10}.

In case of {\bf NGC\,6205} large ranges of light-element abundances have
been found along all   evolutionary phases from the RGB tip down to the MS
\citep[e.\ g.\ ][]{shetrone96, kraft97, briley02}.  NGC\,6205 exhibits a
very extended Na-O and Al-Mg anticorrelation, and a C-N correlation
\citep[see e.\ g. ][ and references therein]{cohen02, sneden04,
smolinski11}.  The RGB is widely spread in colour in the $y$ versus $c_{\rm
y}$ CMD, with at least three components  \citep{grundahl98, yong08,
lardo11},   confirming that this cluster is made of stellar populations
with different nitrogen and carbon content. 

Since the seventies, spectroscopic studies of {\bf NGC\,6397} giants have
shown that the RGB stars exhibit a significant spread in C, N, Na, and O
abundances \citep[e.g.][]{bell79, briley90,  pasquini04,carretta09}, and a
bimodality in sodium and oxygen \citep{lind09, lind11}.  Multi-band {\sl
HST\/} photometry  revealed that the MS of NGC\,6397 splits into two
components made up of $\sim$30\% and $\sim$70\% of MS stars, respectively
\citep{milone12c}, while a bimodal RGB was previously identified by
\citet{grundahl02} and \citet{lind11}.  Multiple stellar populations along
the HB have been investigated by \citet{lovisi12}. The small colour
separation of the two MSs implies that the two stellar populations differ in
helium  by $\Delta$Y$\sim$0.01 \citep{dicriscienzo10a, milone12c}.

%%%%%%%%%%%%%%%%%%%%%%%%%%%%%%%%%%%%%% FIG 7 %%%%%%%%%%%%%%%%%%%%%%%%%%%%%%%%%%%
%%%%%%%%%%%%%%%%%%%%%%%%%%%%%%%%%%%%%%%%%%%%%%%%%%%%%%%%%%%%%%%%%%%%%%%%%%%%%%%

%%%%%%%%%%%%%%%%%%%%%%%%%%%%%%%%%%%%%% FIG 8 %%%%%%%%%%%%%%%%%%%%%%%%%%%%%%%%%%%
%%%%%%%%%%%%%%%%%%%%%%%%%%%%%%%%%%%%%%%%%%%%%%%%%%%%%%%%%%%%%%%%%%%%%%%%%%%%%%%%

%\clearpage

%%%%%%%%%%%%%%%%%%%%%%%%%%%%%%%%%%%%%% FIG 9 %%%%%%%%%%%%%%%%%%%%%%%%%%%%%%%%%%%
%%%%%%%%%%%%%%%%%%%%%%%%%%%%%%%%%%%%%%%%%%%%%%%%%%%%%%%%%%%%%%%%%%%%%%%%%%%%%%%%

%%%%%%%%%%%%%%%%%%%%%%%%%%%%%%%%%%%%% FIG 10 %%%%%%%%%%%%%%%%%%%%%%%%%%%%%%%%%%%
%%%%%%%%%%%%%%%%%%%%%%%%%%%%%%%%%%%%%%%%%%%%%%%%%%%%%%%%%%%%%%%%%%%%%%%%%%%%%%%

%%%%%%%%%%%%%%%%%%%%%%%%%%%%%%%%%%%%% FIG 11 %%%%%%%%%%%%%%%%%%%%%%%%%%%%%%%%%%%
%%%%%%%%%%%%%%%%%%%%%%%%%%%%%%%%%%%%%%%%%%%%%%%%%%%%%%%%%%%%%%%%%%%%%%%%%%%%%%%

Multi-band {\it HST} photometry reveals that the MS, SGB, and RGB of {\bf
NGC\,6752}  splits into three main components \citep{milone13}, in
close analogy with the  three distinct segments along its horizontal branch
stars \citep{momany02,momany04}. These triple sequences are consistent
with three stellar groups: a stellar population with a chemical composition
similar to field halo stars (population a),  a population (c) with enhanced
sodium and nitrogen, depleted carbon and oxygen and enhanced helium
abundance ($\Delta Y \sim$0.03), and a population (b) with an intermediate
(between population a and c) light-element    composition, and slightly
helium enhanced ($\Delta Y \sim$0.01). These components contain $\sim$25\%
(population a), $\sim$45\% (population b), and  $\sim$30\% (population c) of
the cluster stars \citep[see e.g.][for spectroscopic   studies on stellar
populations in NGC\,6752]{norris81,grundahl02,yong05,carretta07,
yong08,carretta12}. 

In {\bf NGC\,7078} CN-CH, C-N, and Na-O anticorrelations have been observed
\citep{sneden97, cohen05, kayser08, pancino10} and also a spread in the slow
neutron-capture elements \citep{sneden97, sneden00, sobeck11}. The RGB is
broadened in the $u-g$ colour from SLOAN photometry \citep{lardo11}.\\  In 
{\bf NGC\,288}, {\bf NGC\,362}, {\bf NGC\,3201}, {\bf NGC\,4590}, {\bf
NGC\,5904},  {\bf NGC\,6218}, {\bf NGC\,6254}, {\bf NGC\,6712}, {\bf
NGC\,6809}, {\bf NGC\,7099}, and {\bf NGC\,6934} multiple stellar populations 
are inferred by star-to-star light-element abundance  variations
\citep{pancino10,cohen05, carretta09,martell08,lee09,kravtsov09, smith86b}.\\  
HST photometry revealed a multiple sequences along the SGB of NGC\,362
\cite{piotto12}. As for NGC\,288 \citep{roh11}, and NGC\,3201 the RGB 
is multimodal or spread  when observed  in appropriate UV bands. In 
{\bf NGC\,6366}, {\bf NGC\,6093}, {\bf NGC\,6541}, and  {\bf NGC\,6981} 
there is no evidence of multiple stellar populations to date, due to lack
of spectroscopic studies of light-element abundances.

\section{Data acquisition, reduction, and standardization}\label{sec:data}

To accomplish the goals listed previously, we started a long-term observing
campaign using facilities in both the northern and southern hemispheres,
namely the Isaac Newton Telescope (INT) operated by the ING in La Palma
(Spain), and the 2.2m MPG/ESO telescope located in La Silla (Chile). Both
telescopes are equipped with comparable large field of view cameras: the WFC
- Wide Field Camera (34$\arcmin$$\times$34$\arcmin$, pixel scale
0.333$\arcsec$/px) and the WFI - Wide  Field Imager
(34$\arcmin$$\times$33$\arcmin$, 0.238$\arcsec$/px), respectively. Both
telescopes have a set of Johnson filters, including the $U$-band one. 

The new data were used to complement the extensive data base maintained by
one of us  \citep[P.B. Stetson, see][]{stetson00}, and in particular to add
the images in the $U$-band for a large number of clusters. So far, the data
have been collected in four different observing runs at the INT, between
August 2011 and August 2012, and one observing run at the 2.2m in February
2012 (complemented by few images collected in service time). We emphasize
that this is a long-term, still ongoing project: the data reduction of the
new data has been completed for a subset of  the targets only, and first
results are presented in the following sections.

The pre-reduction was performed with standard IRAF procedures to correct for
bias, flat-field and fringing in the case of the $I$-band images. The
photometric reduction was performed using DAOPHOT/ALLFRAME \citep{dao,alf}
simultaneously on all the data available for each cluster.  Individual 
PSFs were modeled for each image, sampling the whole area of the chip with 
a large number stars (tens to hundreds), the exact number depending on the 
stellar density of each field. After the ALLFRAME run, resulting magnitudes 
were corrected aperture using DAOGROW \citep{daogrow}, using a constant 
correction. The calibration was performed following \citet{stetson00}. 
So far, artificial stars tests have not been performed.
Table \ref{tab:tab01} presents a
summary of the data relative to the clusters included in  the present paper.
For each target, we indicate R.A. and Dec. (from the compilation by
\citealt{harris10}), and the maximum number of images a given stars has been
measured in.

\section{Photometric selections and Colour-magnitude diagrams}\label{sec:selections}

In this section we describe the procedure followed to obtain the cleanest
possible catalogue for each cluster.  Stars have been selected according to
the photometric error, $\chi{^2}$, and {\itshape sharpness} parameters
provided by ALLFRAME. In some cases we removed the stars located in the
innermost, most crowded regions. The exact selections were optimized for
each cluster.

In the following subsections, we describe the technique adopted to correct
for differential reddening and decontaminate for the presence of foreground
and background Galactic stars and background unresolved galaxies.

	\subsection{Differential reddening}\label{sec:reddening}

We were careful about correcting our photometry for differential reddening
effects by using the procedure described in \citet{milone09}. Briefly, we
first defined the ridge line for the cluster MS and RGB. For each star we
identified the closest neighbours (typically 50), and for each of them we
estimated the amount of shift to the red or the blue side of the fiducial
sequence. This systematic colour offset is indicative of  the local
differential reddening. We applied this method using two CMDs: $V$ {\em vs}
$(U-V)$ and $I$ {\em vs} $(B-I)$, to have two estimates using independent
colour combinations. The final reddening map, presented in Fig.
\ref{fig:redde}  for the case of NGC\,6121 (M4), was obtained as the mean of
these two ones.  The left panel shows the reddening map as a function of the
XY coordinates, according to the colour code displayed at the top of the
plot. In this case, we estimated that the extinction differences are of the
order of $\Delta E(B-V)= \pm$0.05 mag. The central and right panels of
Fig.~\ref{fig:redde} show the $B$ {\em vs} $(B-I)$ CMD of the same cluster
before and after applying the correction for differential reddening.
Clearly, all the features of the CMD become better defined, and the colour
spread significantly decreases, especially in the case of the SGB, RGB and
HB.  Photometry corrected for differential reddening has been used in 
all the diagrams presented in the following.

\subsection{Removing contaminating sources}\label{sec:decontaminations}

To clean the CMDs from both foreground Galactic stars and background
unresolved galaxies we followed a procedure similar to \citet{bono10} and
based on the source position in the ($B-I$) {\em vs} ($U-V$) colour-colour
plane. Figure~\ref{fig:colcol} shows the case of NGC\,6205 and NGC\,6121.
The sources inside the overplotted line are considered bona-fide cluster
stars, while those outside will be rejected from the following analysis. We
stress that this procedure does not work equally well for every cluster. In
fact, in  case of more metal-rich clusters, the RGB gets redder in the $U-V$
colour, making  more difficult to establish the separation from field stars.
This is clearly seen in case of NGC\,6121, where the non-negligible
contamination from bulge stars cannot be completely removed, due to the
superposition of the sequences in these diagrams.

%%%%%%%%%%%%%%%%%%%%%%%%%%%%%%%%%%%%% FIG 12 %%%%%%%%%%%%%%%%%%%%%%%%%%%%%%%%%%%
\begin{centering}
\begin{figure*} 
\includegraphics[width=8.5 cm]{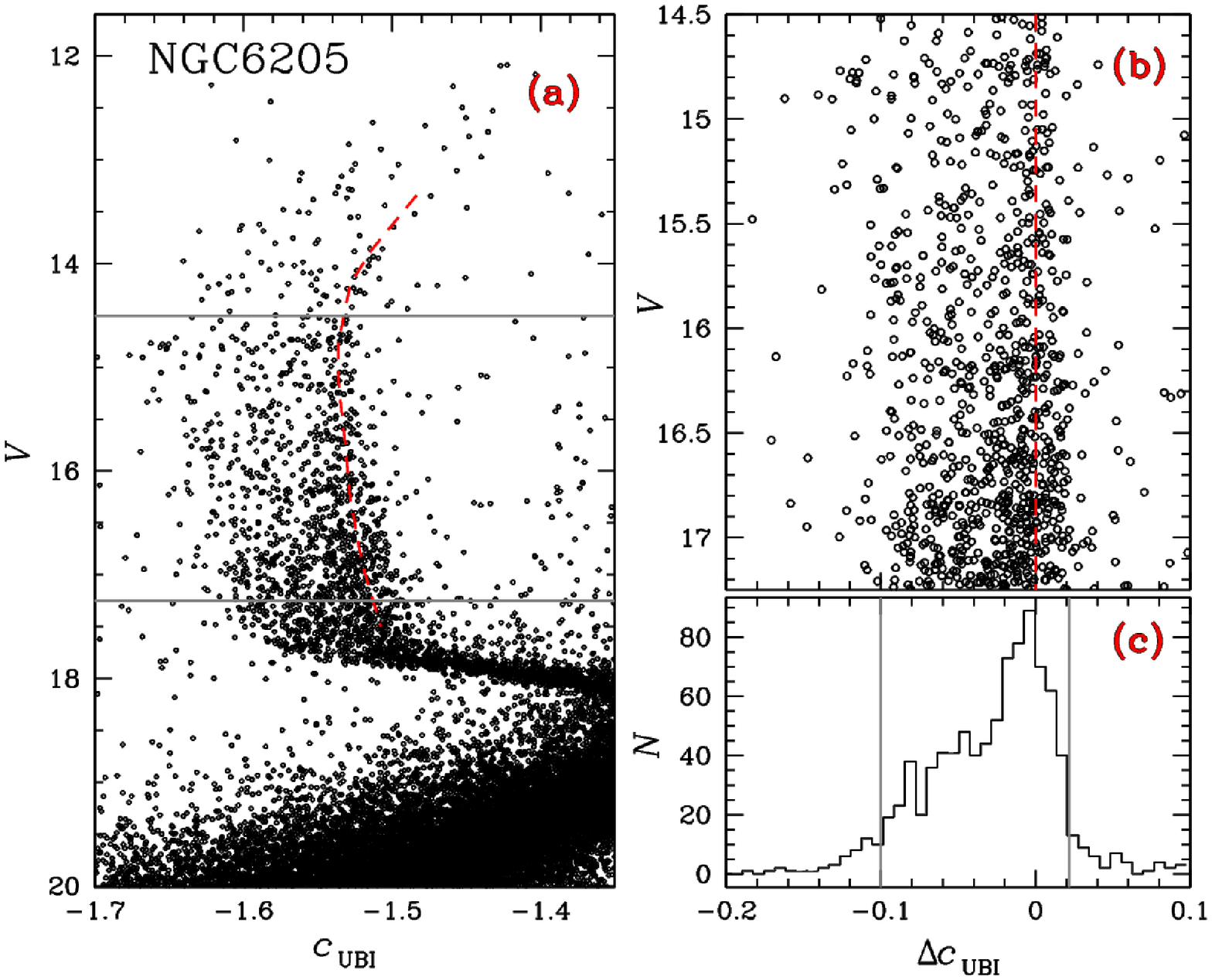}
\caption{Sketch of the
procedure to estimate the RGB width for NGC\, 6205. {\it Panel (a).} $V$-\cubi~ 
diagram zoomed around the RGB. The red-dashed line is the RGB
fiducial line and is drawn by hand. {\it Panel (b).} Normalized $V$-$\Delta$\cubi ~
diagram for RGB stars between the two grey lines of Panel (a). See
text for details. {\it Panel (c).} Histogram of the colour distribution for the RGB stars
displayed in panel (b).} 
\label{fig:histo_proc} 
\end{figure*}
\end{centering}
%%%%%%%%%%%%%%%%%%%%%%%%%%%%%%%%%%%%%%%%%%%%%%%%%%%%%%%%%%%%%%%%%%%%%%%%%%%%%%%%

%%%%%%%%%%%%%%%%%%%%%%%%%%%%%%%%%%%%%% FIG 13 %%%%%%%%%%%%%%%%%%%%%%%%%%%%%%%%%%%
\begin{figure*}
 \includegraphics[width=14cm]{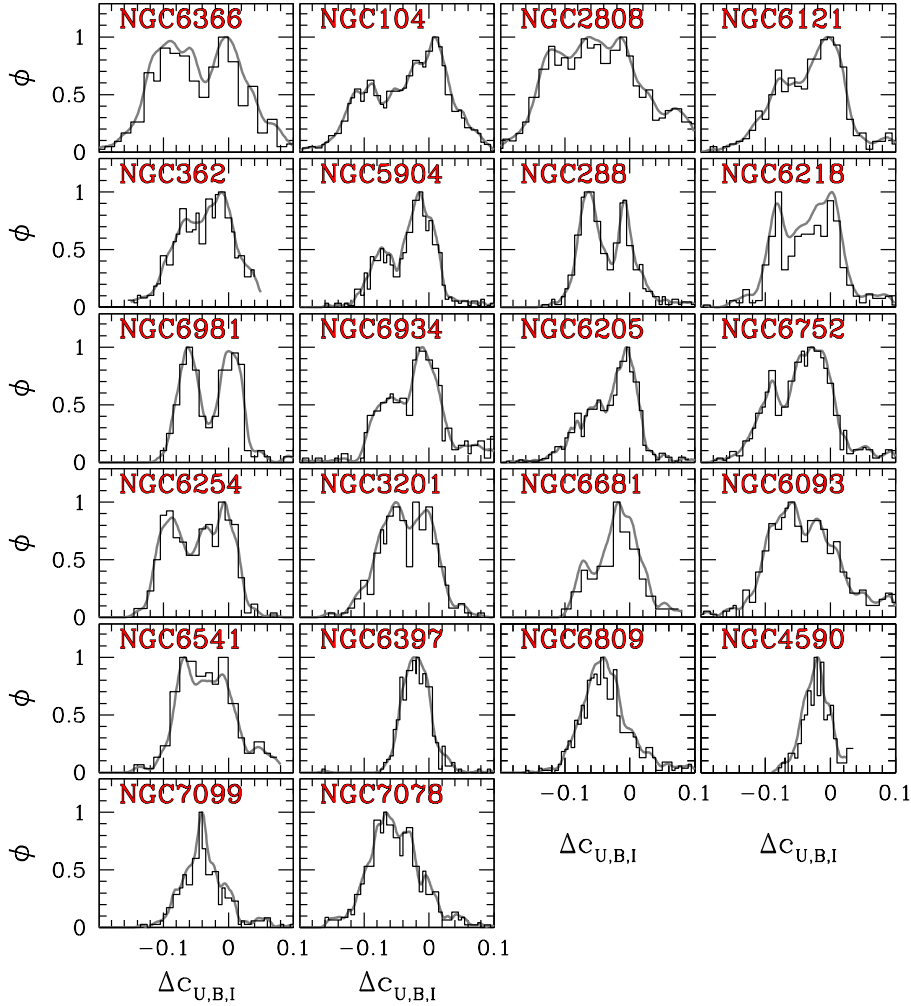}
 \caption{$\Delta$\cubi~ histogram for the 22 clusters analyzed. The broad or complex morphology
 with multiple peaks is common to all globular clusters
 }
 \label{fig:histo}
\end{figure*}
%%%%%%%%%%%%%%%%%%%%%%%%%%%%%%%%%%%%%%%%%%%%%%%%%%%%%%%%%%%%%%%%%%%%%%%%%%%%%%%%

%%%%%%%%%%%%%%%%%%%%%%%%%%%%%%%%%%%%% FIG 14 %%%%%%%%%%%%%%%%%%%%%%%%%%%%%%%%%%%
\begin{centering}
\begin{figure*}
 \includegraphics[width=16cm]{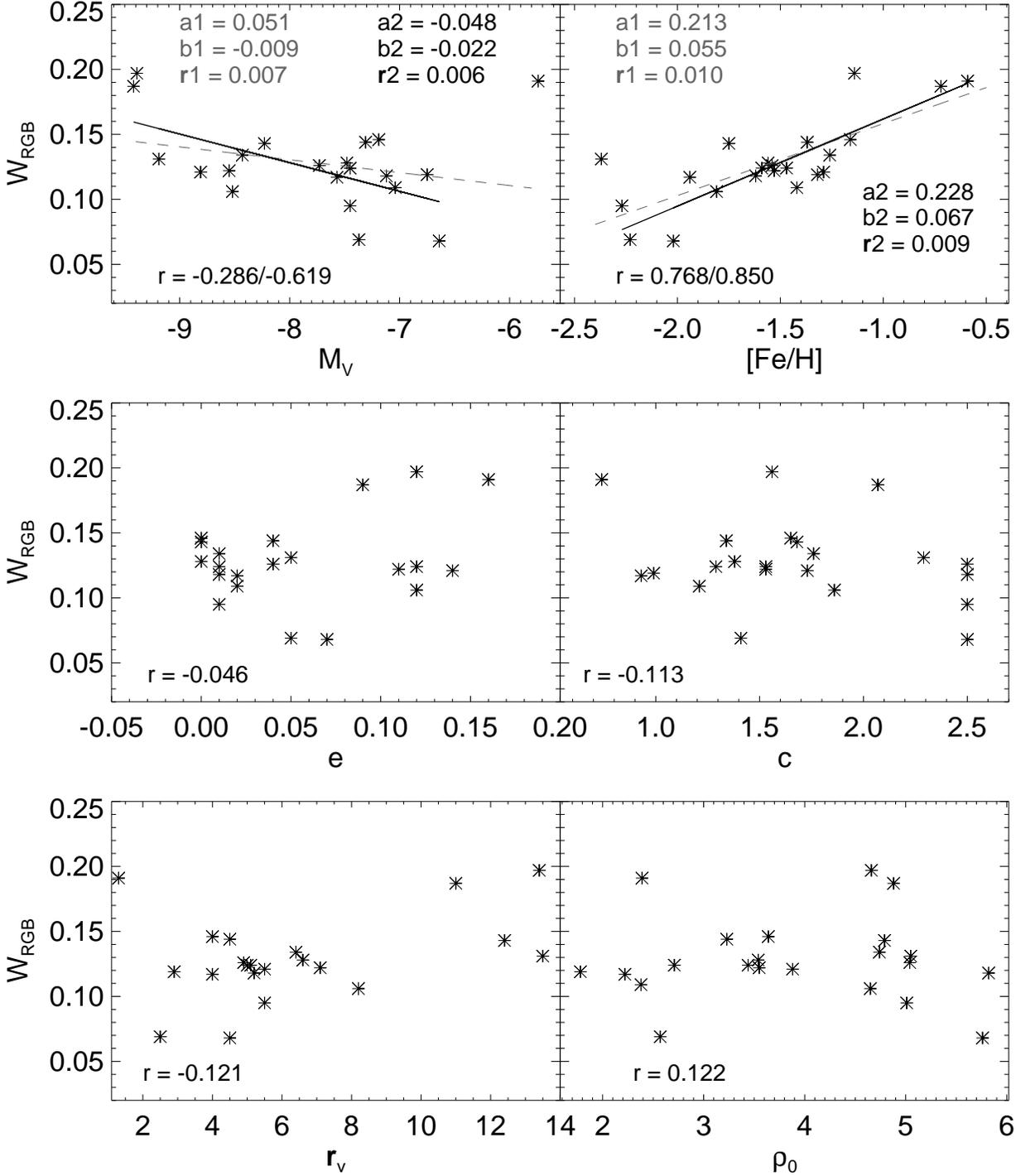}
\vspace{1.cm}
 \caption{The width of the RGB, $W_{\rm RGB}$, is shown as a function of a variety of
 cluster global parameter. In each panel we report the Pearson correlation coefficient.
 In case of $M_V$ and iron content -- shown in the top panels -- we also present a linear
 fit to the sample (dashed gray lines) and the relative coefficients (a1,b1,$\sigma1$). The black lines
 show a similar fit once removed the two outliers, NGC\,6366 in the case of $M_V$ and
 NGC\,7078 in the case of metallicity.}
 \label{fig:width}
\end{figure*}
\end{centering}
%%%%%%%%%%%%%%%%%%%%%%%%%%%%%%%%%%%%%%%%%%%%%%%%%%%%%%%%%%%%%%%%%%%%%%%%%%%%%%%

%%%%%%%%%%%%%%%%%%%%%%%%%%%%%%%%%%%%%% FIG 15 %%%%%%%%%%%%%%%%%%%%%%%%%%%%%%%%%%%
\begin{figure*}
 \includegraphics[width=9cm]{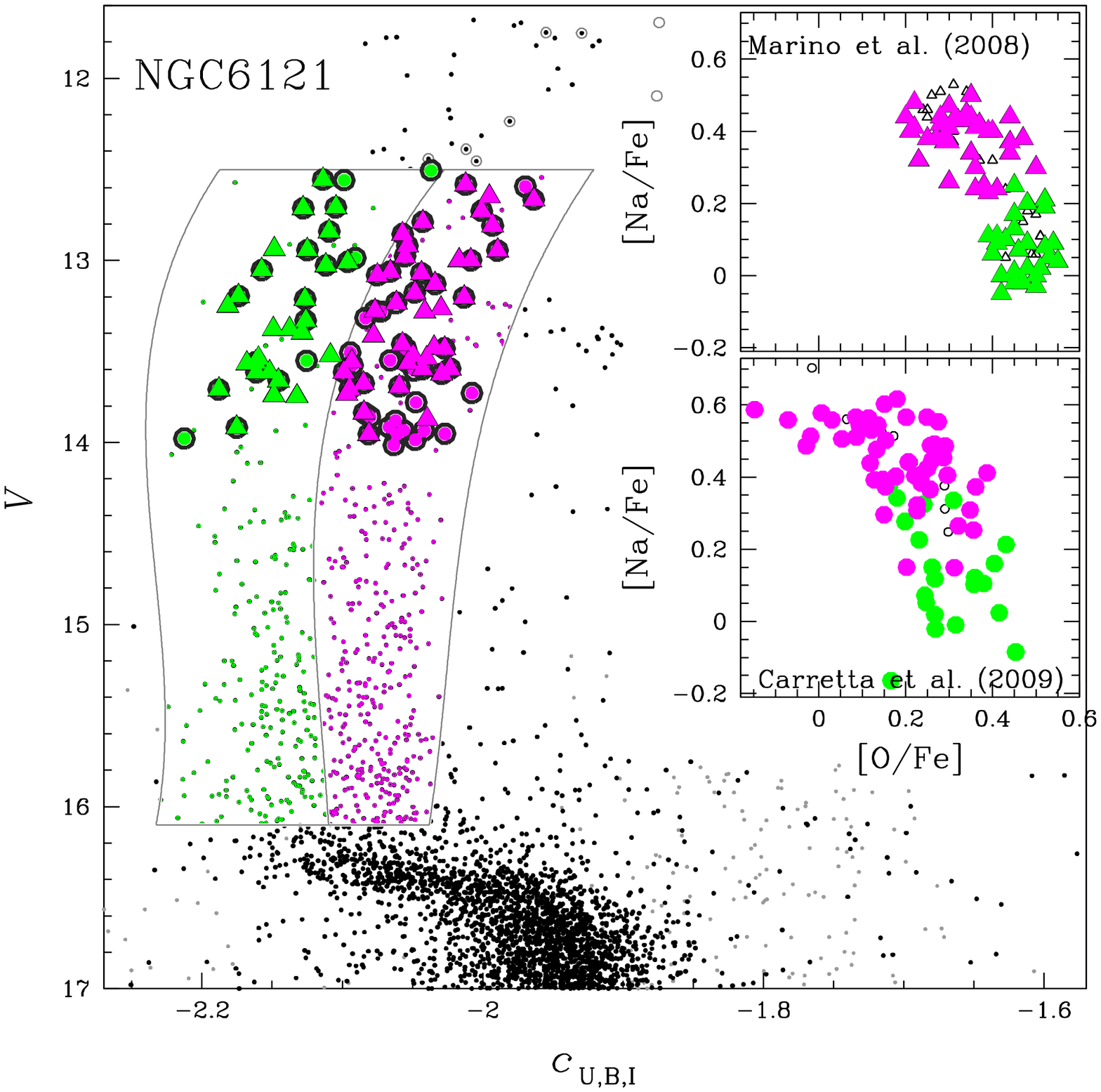}
 \includegraphics[width=9cm]{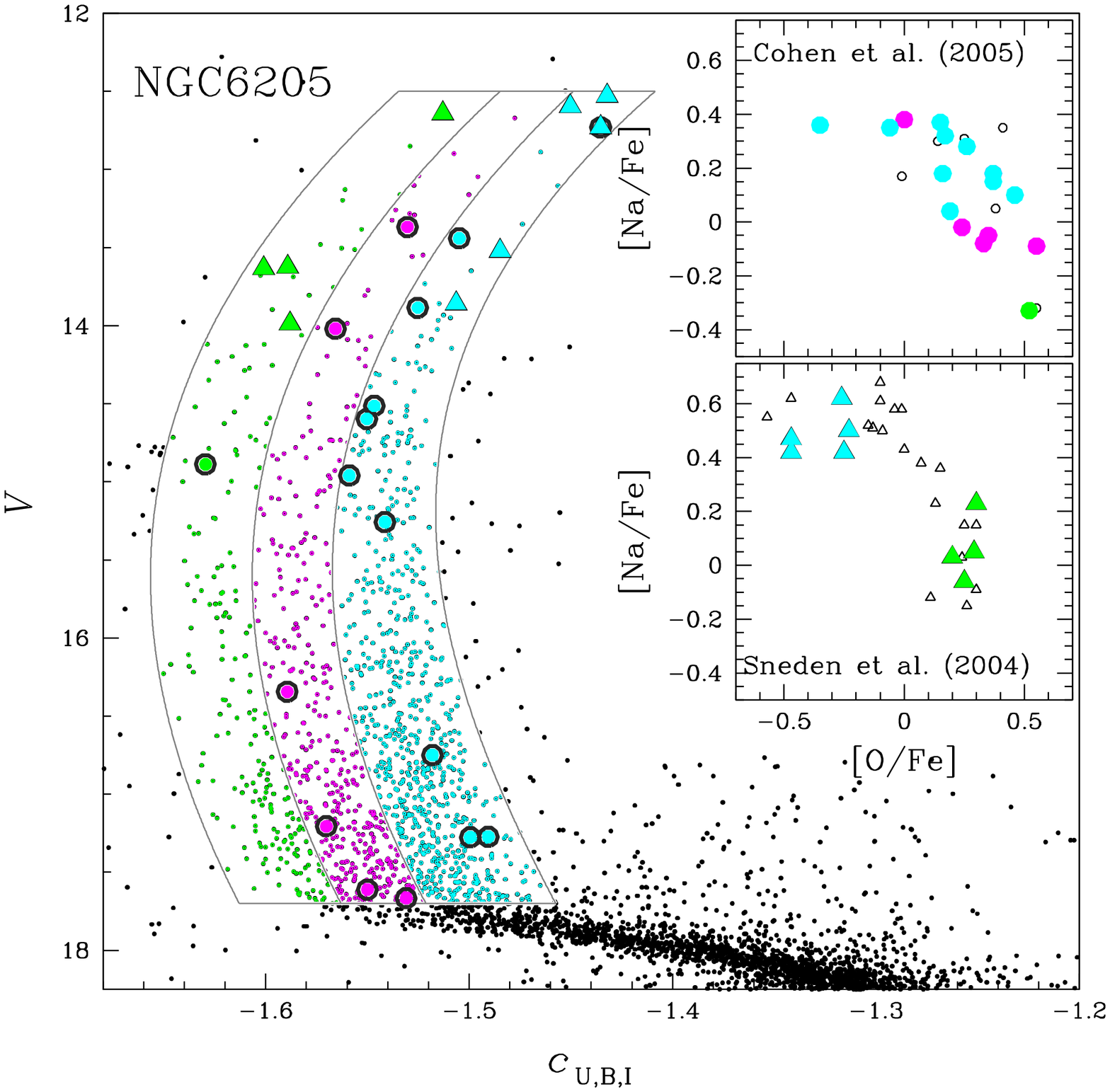}
 \caption{Correlation between the photometric and spectroscopic properties of RGB stars.
 The subpopulations selected in the ($V$,\cubi) plane are found to have different chemical
 content in terms of O and Na abundance.}
 \label{fig:spec}
\end{figure*}
%%%%%%%%%%%%%%%%%%%%%%%%%%%%%%%%%%%%%%%%%%%%%%%%%%%%%%%%%%%%%%%%%%%%%%%%%%%%%%%%

%%%%%%%%%%%%%%%%%%%%%%%%%%%%%%%%%%%%%% FIG 16 %%%%%%%%%%%%%%%%%%%%%%%%%%%%%%%%%%%
\begin{centering}
\begin{figure*}
 \includegraphics[width=15cm]{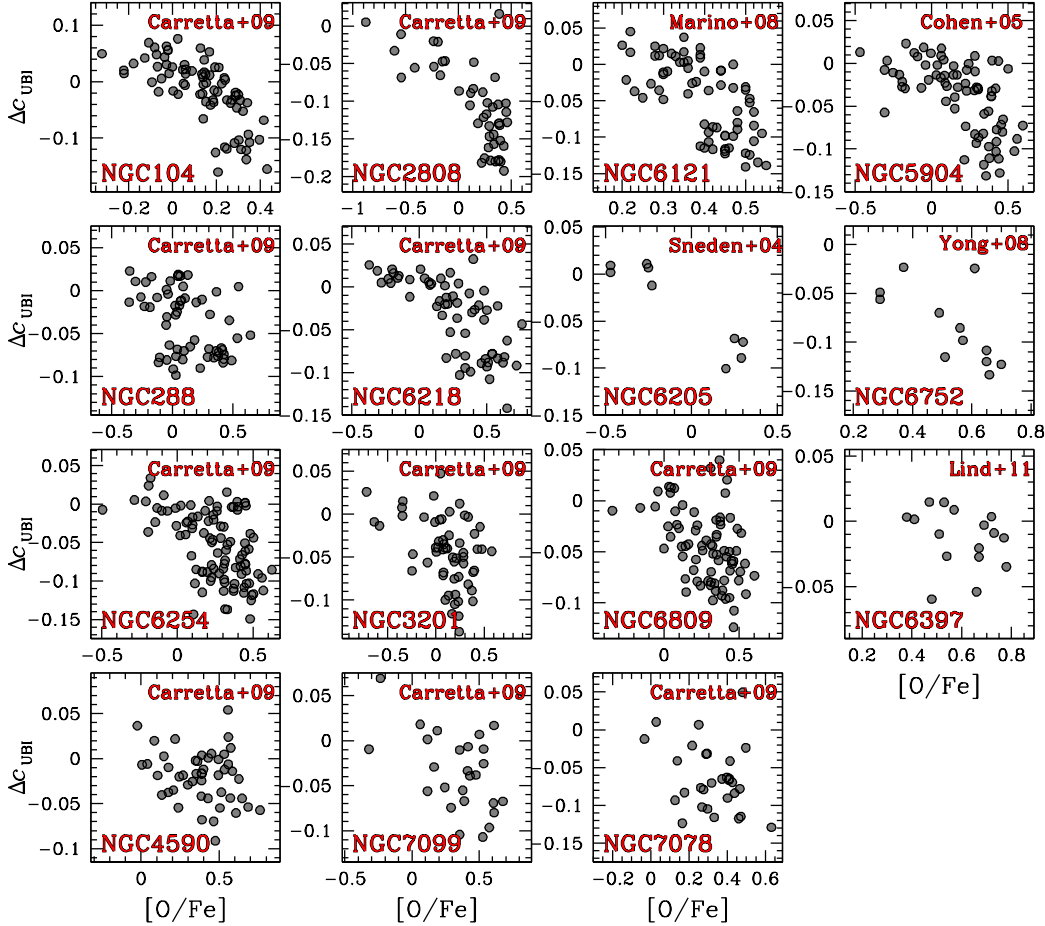}
 \caption{$\Delta$\cubi~ {\em vs} [O/Fe] abundance anticorrelation, found in 
 all the clusters studied.}
 \label{fig:oxygen}
\end{figure*} 
\end{centering}
%%%%%%%%%%%%%%%%%%%%%%%%%%%%%%%%%%%%%%%%%%%%%%%%%%%%%%%%%%%%%%%%%%%%%%%%%%%%%%%%

%%%%%%%%%%%%%%%%%%%%%%%%%%%%%%%%%%%%%% FIG 17 %%%%%%%%%%%%%%%%%%%%%%%%%%%%%%%%%%%
\begin{centering}
\begin{figure*}
 \includegraphics[width=15cm]{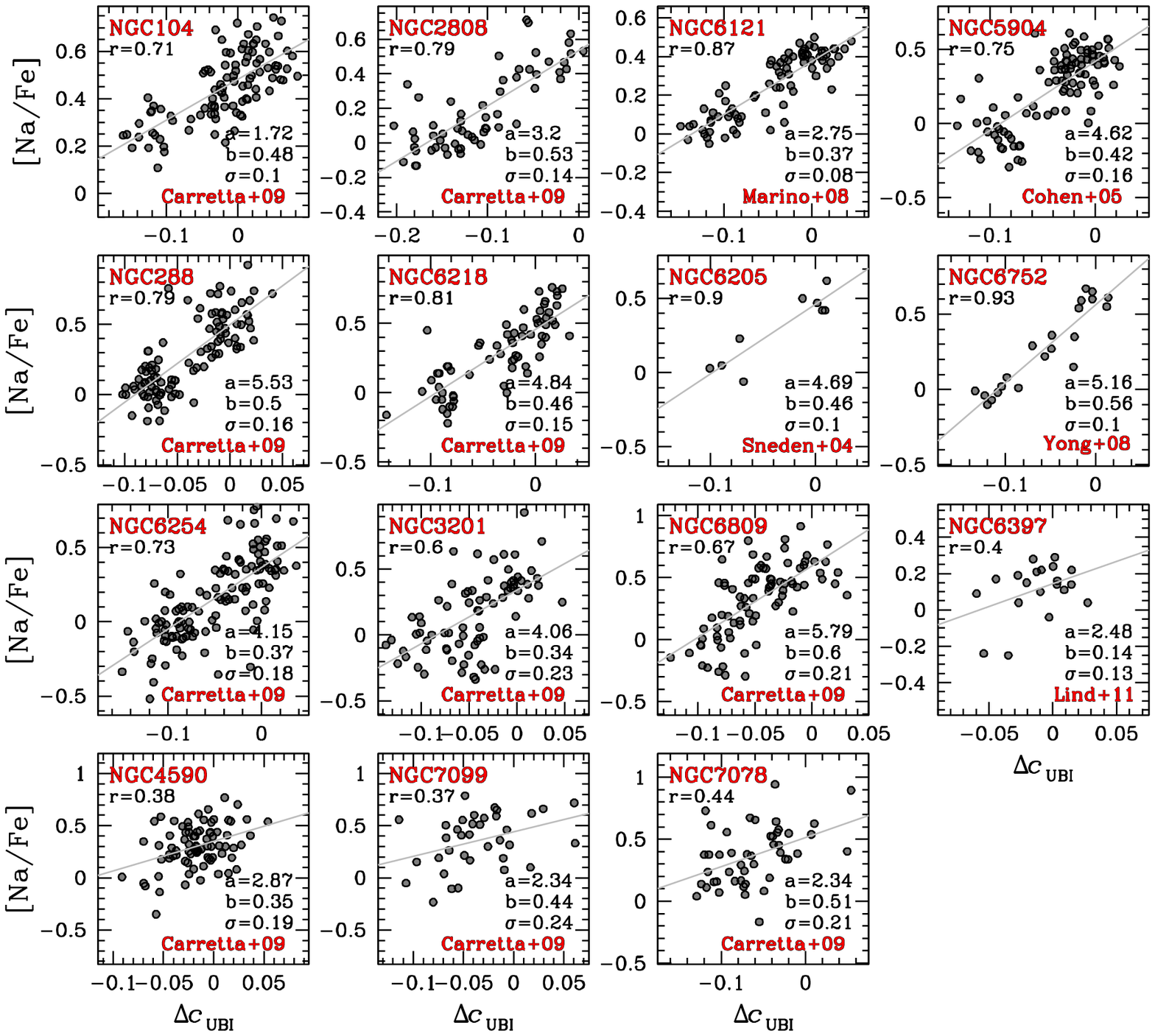}
 \caption{[Na/Fe] {\em vs} $\Delta$\cubi~ correlation, with overplotted the coefficients of
 the linear fits.}
 \label{fig:sodium}
\end{figure*} 
\end{centering}
%%%%%%%%%%%%%%%%%%%%%%%%%%%%%%%%%%%%%%%%%%%%%%%%%%%%%%%%%%%%%%%%%%%%%%%%%%%%%%%%

	\subsection{Colour-Magnitude Diagrams}\label{sec:cmds}
	
Figure~\ref{fig:cmds} shows a number of different CMDs of the clusters
NGC\,6121 (left column) and NGC\,6205  (right), adopted again as a test
case, using different colour combinations. In particular,  the top row
presents the $V$ {\rm vs} ($B-I$) diagrams, while the other panels display  $U$ {\em
vs} ($U-B$), ($U-V$), and ($U-I$), respectively. The figure discloses that
our photometry reaches  $\sim$3 mag below the turn-off in the four bands. The
sources rejected using the colour-colour plane are represented in light
grey,  while the bona-fine cluster stars are in black. We highlighted the
AGB stars in orange, while HB stars are coloured according to the position
relative to the "Momany jump" \citep[][]{momany02} and the "Grundahl jump"
\citep[][]{grundahl99}. In particular, blue, cyan, and green symbols
represent  the stars bluer than the Momany jump, intermediate, and redder
than the Grundahl jump, respectively, while red dots mark the red HB stars.
The selections were performed in  the $V$ {\rm vs} ($B-I$) plane. Note that in case
of M\,4, the red and the blue HB stars  (red and green points, respectively)
are swapped in the ($U-B$) plane. Similarly, the  bluest AGB stars (orange
points) gets bluer  than the reddest HB stars in the same plane. This
behaviour is predicted by theoretical models, and is due to the sensitivity
of $U$ filter to the Balmer jump feature.

\section{The \cubi~ index as a new diagnostic for multiple stellar populations}\label{sec:uband}

	\subsection{Definition}\label{sec:def}

The selected catalogues were employed to look for the evidence of multiple
stellar populations in the observed clusters. Recent studies have shown that
CMDs of globulars  consist of intertwined sequences of two or more stellar
populations, that can be traced continuously from the MS up to the RGB and
the HB \citep{milone12a, milone12b}. Previous studies of stellar populations
in GCs based on Str\"omgrem \citep[e.\ g.\ ][]{grundahl99, yong08},
multi-wavelength {\it HST} and ground-based photometry \citep{marino08, 
bellini10, milone12a, milone12b} have demonstrated that appropriate
combinations of colours and magnitudes can be powerful techniques to
isolate multiple sequences in the CMD of GCs. 

 The ($U-B$) colour is sensitive to light-element variations
\citep{marino08, sbordone11}. These authors have shown that, at least in
some clusters, stars enhanced in N and Na and depleted in O and C, populate
the red part of the RGB in the $U$ versus ($U-B$) CMD, while Na/N-poor and
O/C-rich stars occupy the red RGB region. A visual inspection at these
diagrams reveals that the RGB is  much  broader than expected from
photometric colour errors only, that for RGB stars are typically smaller
than 0.03 mag. 

The ($B-I$) colour is very efficient in disentangling stellar populations
with different helium abundance \citep{piotto07,dicriscienzo11}. Both
theoretical arguments and observations show that  Na/N poor stars are also
helium-poor while stars enhanced in sodium and nitrogen have also higher 
helium abundance. Moreover, helium-rich stars have bluer ($B-I$) but redder
($U-B$) colour than helium-poor ones. Driven by these results, we introduce
here a new index \cubi, defined as $(U-B) - (B-I)$, that maximizes the
separation among stars with different helium and light-elements content.  

In the following, we will use NGC\,6205 and NGC\,6121 to describe the
properties of the \cubi~ index. Figure~\ref{fig:cubi6205} presents the $V$
{\em vs} \cubi~ diagram for both clusters: the left and central panels  show
the entire diagram and a zoom in the RGB region, respectively. In this
plane, the pseudo-CMD appears reversed when compared to a typical CMD:  the
\cubi~ index gets larger (i.e. less negative) for MS stars of increasing 
brightness, while the SGB bends towards more negative colours. Interestingly, 
in the case of NGC\,6121 the red HB crosses the RGB and the blue HB
extends to the larger \cubi~ index, while in the case of NGC\,6205 it is 
completely outside the range of the plot (\cubi ~$> -0.5 mag$, and not shown for clarity.
An homogeneous investigation of the HB morphology will be presented
in a forthcoming paper.

 Contrary to what shown in Fig. \ref{fig:cmds}, the RGB shows a large
colour spread, significantly larger that the photometric error, and multiple
sequences can be recognised in both clusters. This suggests that multiple 
stellar populations are present and that the \cubi~ index is able to identify 
them, by splitting the RGB into multiple components.  In particular, in the
case of NGC\,6121 Figure~\ref{fig:cubi6205} suggests that the RGB is
bimodal in the magnitude interval $\sim 12.5<V<\sim 16.1$. We have thus
drawn by hand the grey lines of Fig.~\ref{fig:cubi6205} (top right panel) to
separate two groups of stars in the magnitude bin where the RGB split is
more evident ($12.5<V<16.1$). These stars are coloured green and magenta,
and named RGBa and RGBb, respectively. Also in the case of NGC\,6205 the RGB
is broadened, with the presence of at least three components. Similarly to what
done for the case of NGC\,6121, in Fig.~\ref{fig:cubi6205} (bottom right) we
have selected three  groups of stars along the RGB of NGC\,6205 (RGBa,b,c),
that have been plotted  with  green, magenta, and cyan colours.

The selected populations help to understand why the \cubi~ is a good  colour
combination to split the different RGBs that otherwise are photometrically
degenerate.  This is shown in Fig. \ref{fig:works}. The top three panels
show the ridge  lines of the two sequences detected in NGC\,6121, in three different CMDs ($V$ {\em vs} $(U-B)$, ($B-V$), ($B-I$)). The same is shown for the three sequences of NGC\,6205 in the central row. 
The crucial point is that the {\itshape relative} position of different ridge lines
changes depending on the colour index adopted. This is clearly seen for NGC\,6121,
where the green line is bluer than the magenta line in $U-B$, but the opposite
occurs in the other two planes. A similar trend is seen in NGC\,6205, where
the bluest sequence in $U-B$ (the green line again) is the reddest in the other
planes. To quantify this effect, we calculate the colour difference between
the ridge lines, at a fixed magnitude level roughly corresponding to a point
two $V$ magnitudes brighter than the MSTO. In particular, we fixed $V = 14.82$ 
mag for NGC\,6121 and $V$=16.63 mag in the  case of NGC\,6205. We calculated 
the colour differences with respect of the green line: the values are 
reported in each panel, and we plotted them in the bottom panels of the figure
as a function of the central wavelength of the adopted filter. Namely, we 
show the $\Delta$($U-B$), $\Delta$($B-V$), and $\Delta$($B-I$) differences. The evidence 
that the ridge lines corresponding two the distinct sub-populations change 
their relative location in the various observational planes is reflected by the 
fact that the colour difference changes sign. From this plot one can immediately conclude 
that the colour difference $(U-B) - (B-I)$  is that one that {\itshape maximizes}
the split between the different RGB sequences. Also note that the
\cubv ~= $(U-B) - (B-V)$ combination can work, but less efficiently than the 
\cubi~ index. We shown an example of this in the Appendix.

We note that the \cubi~ index provides a pseudo-CMD morphologically similar
to the $cy$ index introduced by \citet{yong08a}, that combines three
Str\"oemgren colours. The advantage of the \cubi~ index is, from the
observational point of view, that it is based on broad-band photometry, more
accessible and less time consuming than the narrow-band technique. As shown
by \citet{yong08}, the $cy$ index is a good tracer of the abundance of N  in
RGB stars. As for the case of $cy$, the reason for the efficiency of the
\cubi~ index to split  subpopulations along the RGB has to be searched in the
spectral properties of the different populations. This will be discussed in
\S \ref{sec:spec}. However, we stress  here that an efficient photometric
identification of different stellar populations along the RGB is  the first
step to guide future spectroscopic campaigns.

%%%%%%%%%%%%%%%%%%%%%%%%%%%%%%%%%%%%%% FIG 18 %%%%%%%%%%%%%%%%%%%%%%%%%%%%%%%%%%%
\begin{centering}
\begin{figure*}
 \includegraphics[width=18cm]{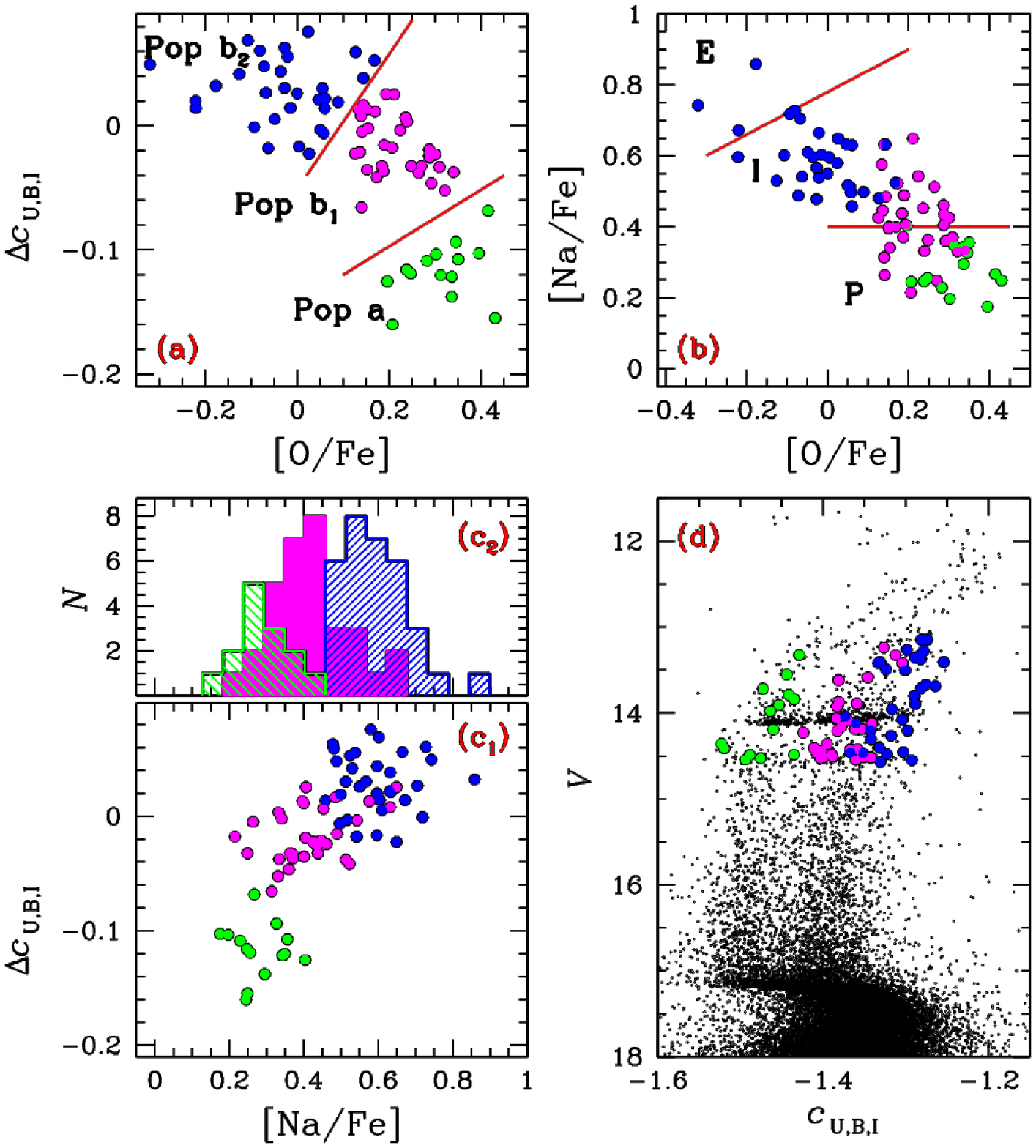}
 \caption{
 \textit{Panel a}: Reproduction of the $\Delta$\cubi-O anticorrelation 
 of Fig.~16 for NGC\,104. Green, magenta, and blue circle mark the three populations 
 $a$, $b_{1}$, and $b_{2}$. \textit{Panel b}: Na-O anticorrelation. Red segments 
 separate the three groups of P, I, and E stars defined by \citet{carretta09}. 
 \textit{Panel $c_1$}: 
 Na-$\Delta$\cubi~ correlation. $V$ {\em vs} \cubi~ diagram.
}
 \label{fig:47trgb}
\end{figure*}
\end{centering}
%%%%%%%%%%%%%%%%%%%%%%%%%%%%%%%%%%%%%%%%%%%%%%%%%%%%%%%%%%%%%%%%%%%%%%%%%%%%%%%%

%%%%%%%%%%%%%%%%%%%%%%%%%%%%%%%%%%%%%% FIG 19 %%%%%%%%%%%%%%%%%%%%%%%%%%%%%%%%%%%
\begin{centering}
\begin{figure*}
 \includegraphics[width=16cm]{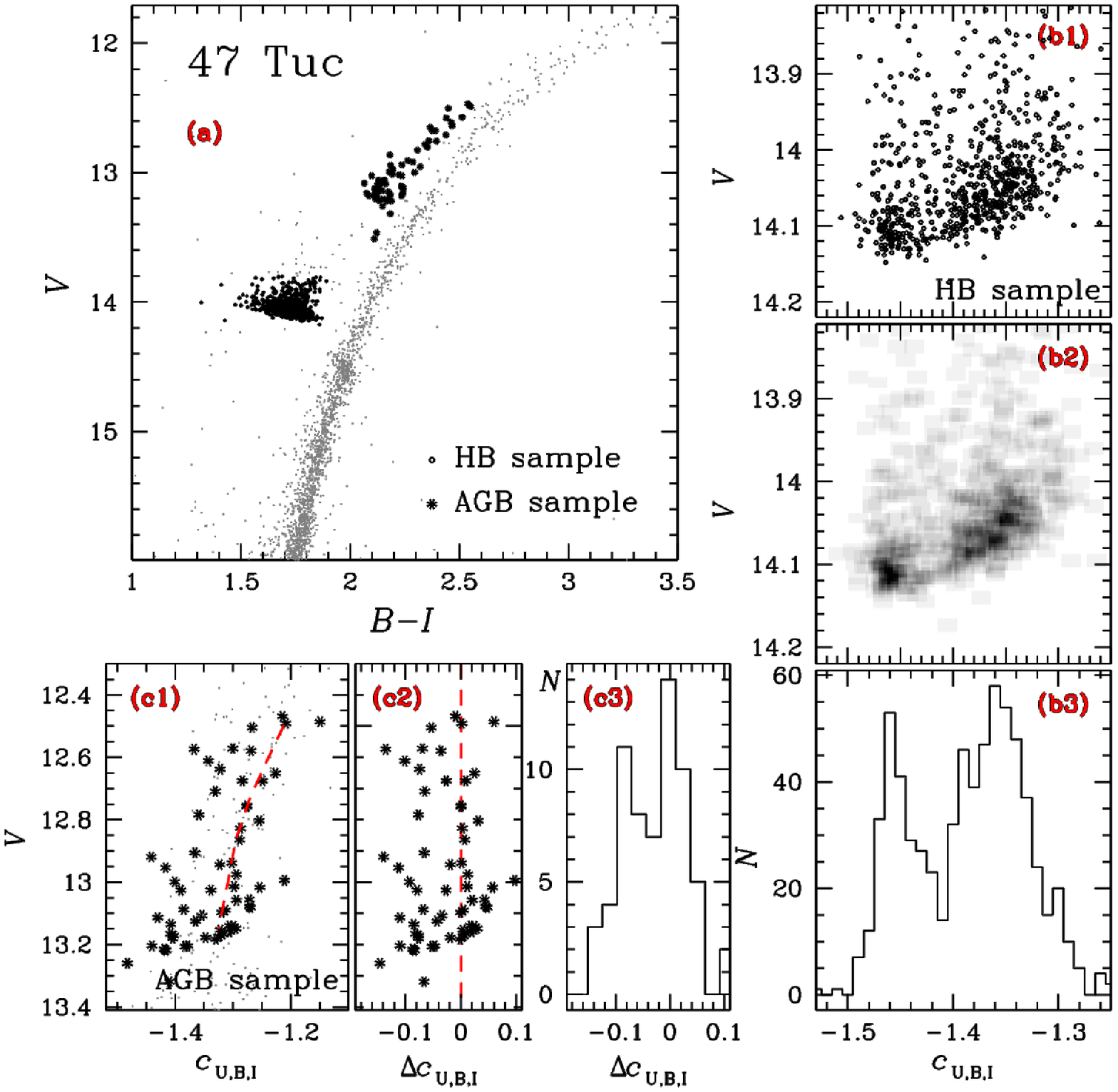}
 \caption{
\textit{Panel a:} $V$ {\em vs} $B-I$ CMD of NGC\,104 (47 Tuc). HB and AGB stars are 
represented with black circles and asterisks, respectively.  
\textit{Panels b:} Zoom of the $V$ versus \cubi~ diagram (panel b1) and Hess 
diagram (panel b2) around the HB. The histogram of the $\Delta$ \cubi~ distribution of HB stars 
is shown in the panel b3.
\textit{Panels c:} $V$ versus \cubi~ diagram of 47 Tuc zoomed around the AGB; 
the red dashed line is a fiducial line drawn by hand through the AGB (panel c1).
Panel c2 shows the \lq{verticalized}\rq $V$ versus $\Delta$ \cubi~ diagram obtained 
by subtracting from the \cubi~ value of each star the corresponding value 
of the fiducial at the same $V$ magnitude. Panel c3 shows the histogram of the 
$\Delta$ \cubi~ distribution for AGB stars.
 }
 \label{fig:HBAGB47tuc}
\end{figure*}
\end{centering}
%%%%%%%%%%%%%%%%%%%%%%%%%%%%%%%%%%%%%%%%%%%%%%%%%%%%%%%%%%%%%%%%%%%%%%%%%%%%%%%%

	\subsection{The \cubi~ index: multiple RGBs in 22 GCs}\label{sec:cubicluster}

The $M_V$ {\em vs} \cubi~ diagrams for a sample of 22 clusters observed in
this survey are shown in Fig.~\ref{fig:cubi}--\ref{fig:cubi6}\footnote{The
\cubv~ diagram for NGC\,6712 is presented in the Appendix.}.  To better
compare the different clusters, we have adopted the values of distance and
average reddening from \citep[][updated as in 2010]{harris96} and shifted
the diagrams to absolute magnitudes.

Clearly, the RGBs present large colour spreads, and in most cases multiple
sequences are evident. This strongly suggests that multiple stellar
populations are a common feature  in globular clusters.  Note that the clusters are sorted for
decreasing metallicity in Fig.~\ref{fig:cubi}--\ref{fig:cubi6}. By comparing
the pseudo-CMDs   one can disclose interesting features. First of all, the
slope of the RGBs in this diagram is strongly sensitive to metallicity. In
fact, while the \cubi~ index of the RGBs in the most metal-rich clusters
([Fe/H]$\gtrsim-1.3$) gets less negative for brighter magnitudes, the more
metal-poor the cluster, the more negative in index for the brightest stars. 
For metallicities close to [Fe/H]$\sim-1.3$ the RGBs are almost vertical.

Qualitatively, not only the slope but also the morphology of the RGB change
from cluster to cluster. In particular, some cluster display a RGB dominated
by two sub-branches, the best examples  being NGC\,6981 and NGC\,5904, while
others are characterized by a smoother distribution of stars (see NGC\,2808,
NGC\,6205). Nevertheless, continuous does not imply homogeneous
distribution.  Overdensities and substructures  are evident, suggesting that
in all clusters there are dominant subpopulations, but in none of them the
RGB appears neatly separated into several well defined sequences.  The
presence of a few stars in between might be due to photometric error, but we
cannot exclude that they belong to tails of the distributions, with
intrinsic intermediate properties.

	\subsection{Dependence on the clusters'properties}\label{sec:properties}
	
To further investigate the influence of metallicity and other cluster
parameters on the multimodal RGBs, we have adopted the procedure illustrated
in Fig.~\ref{fig:histo_proc}  for NGC\,6205. Panel (a) is a reproduction of
the Right panels of Fig.~\ref{fig:cubi6205} showing the $V$ versus \cubi ~
diagram zoomed around the RGB, where we have superimposed the red-dashed
fiducial line, that is drawn by hand. The two gray lines are located 1.5 and
4.5 mag  (in the $V$ band) above the MS turn-off and delimit the region
where the RGB multimodality is better visible. Only stars in this magnitude
intervals are used in the following analysis. The normalized $V$ versus
$\Delta$\cubi~ diagram is plotted in panel (b), and is obtained by
subtracting from the \cubi~ colour index of each stars the corresponding
value of the fiducial at the same $V$ mag. Panel (c) shows the $\Delta$\cubi ~
histogram distribution. The histogram distributions  for all the clusters
studied in this paper are plotted in Fig.~\ref{fig:histo} for stars in the
same magnitude interval. Superimposed to each of them, we plot the  kernel 
density distribution, obtained using a gaussian kernel.
The \cubi~ distribution of RGB stars confirms the
multimodality detected in the pseudo-CMDs. NGC\,6366, NGC\,104, NGC\,5904, 
NGC\,6121, NGC\,6254, NGC\,288, and NGC\,6981 have clearly bimodal RGBs, while
NGC\,2808,  NGC\,6205, NGC\,6218, and NGC\,6254, present some hint of three
components. 

Starting from the normalized colour histogram previously defined, we defined
the  width of the RGB, $W_{\rm RGB}$, as the colour extension of the
histogram after    removing the 5\% bluest and reddest stars (as an example,
we see the two vertical lines in panel (c) of
Fig.~\ref{fig:histo_proc})\footnote{The outliers have been rejected  to
minimize the effect of field-star contamination and objects with large
photometric  errors on the estimate of $W_{\rm RGB}$.}. The value of the
$W_{\rm RGB}$ index, summarized in  Tab. \ref{tab:tab02}, spans a  limited
range, from 0.06 to 0.20 mag, with mean value $<W_{\rm RGB}>=0.13\pm0.03$ mag.
We searched for possible correlations between the RGB width $W_{\rm RGB}$ and
the global  clusters'parameters. We adopted the compilation by
\citet{harris10},  and calculated the Pearson linear coefficient with a
number of different parameters. Some examples are shown in
Fig.~\ref{fig:width}. The three rows respectively show, from top to bottom,
the  trend of $W_{\rm RGB}$ as a function of cluster integrated absolute visual
magnitude $M_{V}$ and the  iron content, the ellipticity and the
concentration, the internal  velocity dispersion and the central density. 
The Pearson coefficient is reported in each panel. In general, correlations
seem to be very weak. The only two exceptions are the integrated absolute
magnitude and the metallicity. Using the full set of clusters, we derive
correlation coefficients {\itshape r}=$-$0.286 and 0.768, respectively. In
both cases, plot the linear fit together with the data points  (dashed grey
lines) and report the fit coefficients (zero point a1, slope b1, and
dispersion $\sigma1$). Note that in both plots there are outliers with
respect of the general trend. In the case of $M_V$, $W_{\rm RGB}$ seems to
decrease for increasing (fainter) magnitudes, but the faintest cluster in
the sample, NGC\,6366, is an obvious outlier. If we remove it, the
correlation improves from $-$0.286 to $-$0.619. The fit without this cluster
gives a steeper  relation (black line, with a2, b2, $\sigma2$ coefficients).
Note that NGC\,6366 is a bulge cluster,  and we cannot exclude that the low
luminosity is due to severe mass loss during its orbital motion.

A similar situation occurs for the dependence on the metallicity. While
there is a general trend of decreasing $W_{\rm RGB}$ for decreasing 
metallicity, the most metal-poor cluster, NGC\,7078, presents a colour
spread larger than expected according to that relation. Removing this
object, the correlation increases from 0.768 to 0.850, resulting again in a
steeper relation (black line). A possible explanation  is that the \cubi ~
colour becomes less sensitive to the  colour spread of the RGB in this
metal-poor regime. This could be due to the fact that more metal-poor RGBs
are also bluer and hotter, and therefore the spectral features responsible
for the colour spread might be weaker.  However, it is worth noting that a
similar anomalous behaviour of NGC\,7078 was found by \citet{pancino10},
when considering the CN-CH anticorrelations as a function of metallicity for
15 clusters. 

\begin{table*}
  \caption{List of the derived RGB width, $W_{\rm RGB}$, and of the slope of the $\Delta$\cubi~ 
  {\em vs} [Na/Fe] correlation, $b$.\label{tab:tab02}}
  \begin{tabular}{l|c|c|c|}
  \hline 
 Cluster     &  $W_{\rm RGB}$  & \multicolumn{2}{|c|}{$\Delta$\cubi~ {\em vs} [Na/Fe]}         \\
             &             & $b$ [mag]  & r          \\
 \hline
NGC\,104     &    0.19     &  0.48      &   0.71     \\
NGC\,288     &    0.14     &  0.50      &   0.79     \\
NGC\,362     &    0.13     &   ---      &   ---      \\
NGC\,2808    &    0.20     &  0.53      &   0.79     \\
NGC\,3201    &    0.10     &  0.34      &   0.60     \\
NGC\,4590    &    0.07     &  0.35      &   0.38     \\
NGC\,5904    &    0.12     &  0.42      &   0.75     \\
NGC\,6093    &    0.14     &  ---       &   ---      \\
NGC\,6121    &    0.14     &  0.37      &   0.87     \\
NGC\,6205    &    0.12     &  0.46      &   0.90     \\
NGC\,6218    &    0.13     &  0.46      &   0.81     \\
NGC\,6254    &    0.13     &  0.37      &   0.73     \\
NGC\,6366    &    0.19     &   ---      &   ---      \\
NGC\,6397    &    0.07     &  0.14      &   0.40     \\
NGC\,6541    &    0.11     &   ---      &   ---      \\
NGC\,6681    &    0.12     &   ---      &   ---      \\
NGC\,6752    &    0.10     &  0.56      &   0.93     \\
NGC\,6809    &    0.12     &  0.60      &   0.67     \\
NGC\,6934    &    0.12     &   ---      &   ---      \\
NGC\,6981    &    0.11     &   ---      &   ---      \\
NGC\,7078    &    0.13     &  0.51      &   0.44     \\
NGC\,7099    &    0.10     &  0.44      &   0.37     \\
\hline 
\end{tabular}
\end{table*}

Interestingly, we stress that solid evidence supports the existence of 
multiple populations in both NGC\,6397 (the cluster with the smallest
$W_{\rm RGB}$, and NGC\,7078, coming from both spectroscopic \citep{lind11} and
photometric \citep{milone12b} investigations. In particular, NGC\,6397
presents small variation  of light elements, in particular Na and O. This
might also explain the limited colour spread of the RGB we reported here.
Interestingly, when comparing the RGB morphology in the  $V$ {\em vs} \cubi ~
plots, one could speculate that the small colour spread observed in
NGC\,6397 is due to the {\it absence} of the red RGB, which is populated by
the Na-rich, O-poor population. This is in nice agreement with the
spectroscopic measurements, suggesting that the \cubi~ colour can be not only
an effective instrument to detect multiple populations, but also to
constrain the relative importance of different sub-populations in a cluster.

\section{Discussion: the chemical properties of multiple sequences}\label{sec:spec}

To demonstrate the effectiveness of the \cubi~ index, in the following we
will take advantage of the measurements of chemical abundance available in
the literature for NGC\,6121 and NGC\,6205, to get information on the
chemical composition of their multimodal RGBs.

Figure~\ref{fig:spec} is a reproduction of the right panels of Figure
\ref{fig:cubi6205}, zooming on the RGB of NGC\,6121 (top) and NGC\,6205
(bottom) in the ${\it V}$ {\em vs} \cubi~ plane. The colour code is the same,
as well as the definition of RGBa,b and RGBa,b,c for the two clusters. In
case of NGC\,6121, stars with available spectroscopic measurements of Na and
O from \citet{marino08} and \citet{carretta09} are plotted as circles and
triangles respectively, while the Na-O anticorrelation determined by these
authors is shown in the two insets. A visual inspection  immediately reveals
that stars from the RGBa and RGBb are clustered around two distinct values
of Na and O. The RGBa has a chemical composition similar to that of halo
stars, and the RGBb is populated by  O-poor and Na-rich stars. Similar
conclusions can be also inferred from the lower inset, showing the Na-O
correlation from \citet{carretta09}, even if the   separation between RGBa
and RGBb stars is less clear, likely due to the poorer accuracy of this data
set.

In the case of NGC\,6205, we identified three RGB components. The Na-O
anticorrelations for stars of this cluster is available from
\citet{cohen05} and \citet{sneden04}, and is plotted in the  inset. Stars
with both spectroscopic and photometric measurements are marked with
coloured circles and triangles. We note that the three groups of RGB stars
have, on average, different abundances of Na and O. RGBa stars are Na-poor
and O-rich, RGBc stars are Na-enhanced and  O-depleted, and RGBb have
intermediate properties. We emphasize here that the choice of the two or
three groups of stars is  not intended to demonstrate that these groups
correspond to discrete stellar populations. The intent here is to
investigate the connection between multimodal RGB and chemical composition. 

Figures~\ref{fig:oxygen} and ~\ref{fig:sodium} display $\Delta$\cubi~ versus
[O/Fe] and the [Na/Fe] versus $\Delta$\cubi~ for the seventeen clusters with
available spectroscopic measurements. Note that for some clusters various
spectroscopic analysis are available. In these cases we only report the
source which provides the smallest scatter in the [Na/Fe] versus
$\Delta$\cubi~ relation. Fig.~\ref{fig:oxygen} shows a clear anticorrelation
between $\Delta$\cubi~ and [O/Fe].  Remarkably, a similar relation between 
the Str\"omgren index $cy$ and the nitrogen abundance has been observed for
RGB stars of NGC\,6752 by  \citet{yong08a}.

Interestingly,  there appears to be a linear correlation 
[Na/Fe]=a$\Delta$\cubi+b. The values of a and b change from one cluster to
the other and are shown in Fig.~\ref{fig:sodium}, together with the scatter
$\sigma$ around the fit.  Apparently, the slope of the [Na/Fe] versus
$\Delta$\cubi~ relation is different in the different clusters, and the
values are reported in Tab.~\ref{tab:tab02}

Some hint of $\Delta$\cubi-Na correlation and $\Delta$\cubi-O
anticorrelation is visible also in NGC\, 6397, despite the small \cubi ~
broadening observed in this cluster. This supports the presence of multiple
stellar generations. Overall, these results imply that the \cubi~ index can
be used to trace the overall distribution of sodium, and oxygen in GCs.  

	\subsection{New hints on the stellar populations of NGC\,104}\label{sec:104}

An accurate analysis of the stellar populations in each individual GC is
beyond the scope   of this paper. The study of NGC\,104 (47 Tuc) presented
in this section is only meant to emphasize, again,   the effectiveness of
the \cubi~ index. As already mentioned in Sect.~\ref{sec:targets}, a number
of recent studies have focused  on the multiple stellar populations in this
cluster, including works based on {\it HST}  photometry and high-resolution
VLT spectroscopy. In the following, we show that the ground-based photometry
of the SUMO dataset can provide  a view on the stellar populations in
NGC\,104 that is even more complex than that given by {\it HST}.

To do this, we start the discussion from the RGB, where  {\it HST} and
ground-based photometry  revealed two distinct sequences that correspond to
two stellar populations \citep{milone12d}. The less-populated component
(population a) contains about 30\% of the stars and has a chemical 
composition similar to the halo-field stars. The most-populated one
(population b) includes the  remaining $\sim$70\% of stars, it is enhanced
in nitrogen and helium, and depleted in carbon and oxygen.

The $\Delta$\cubi-[O/Fe] anticorrelation of Fig.~\ref{fig:47trgb}a confirms
the presence  of two stellar groups.  A population of O-rich stars with low
$\Delta$\cubi~ values,  and a population of O-poor stars with large
$\Delta$\cubi ~. These two stellar groups  correspond to the populations $a$
and $b$ identified by \citet{milone12d}. Interestingly, while stars
belonging to population $a$ are fairly homogeneous in [O/Fe], 
population-$b$ stars span a wide range of about 0.6 dex in oxygen abundance.
Hence we have  further arbitrarily divided population $b$ into two
subgroups, denoted as $b_{1}$ and $b_{2}$.  Population $a$-, $b_{1}$-, and
$b_{2}$-stars are displayed in Fig.~\ref{fig:47trgb}a with green,  magenta,
and blue colour, respectively. The Na and O estimates come from
\citet{carretta09}.

These authors suggested a way to separate the cluster multiple stellar
populations on the  basis of their position in the Na-O plane. They defined
as primordial (P) component all stars with [Na/Fe] ratio in the range
between  $\rm [Na/Fe]_{\rm min}$ and $\rm [Na/Fe]_{\rm min}+$0.3. Here $\rm
[Na/Fe]_{\rm min}$ is  the minimum value of the ratio [Na/Fe] ratio
estimated by eye.  The remaining stars are all considered
second-population(s) stars, and have been further  divided into two groups.
Stars with ratio [O/Na]$>-$0.9 dex belong to the intermediate (I) 
population, while those with [O/Na]$<-$0.9 dex are defined as extreme (E)
population.  E, I, and P groups are separated by red lines in the [Na/Fe]
{\em vs} [O/Fe] plane shown in Fig.~\ref{fig:47trgb}b.  The \ P component
includes all the stars in population $a$, but is strongly contaminated by  
population $b_{1}$ stars. Similarly, the I component contains a mix of
population $b_{1}$ and  population $b_{2}$ stars.

An effective way  of testing the presence of a real oxygen spread amongst
population $b$ stars,  makes use of a comparison of independent measurements
of different quantities for the same stars, like $\Delta$\cubi,  [Na/Fe],
and [O/Fe]. Specifically, if the broadening is intrinsic, population $a$-,
$b_{1}$-,  and $b_{2}$-stars, will also differ in their sodium abundance. 
But if the broadening is  entirely due to observational errors, the stars in
the three stellar populations have the  same probability of being Na-rich or
Na-poor.   Panel c$_1$ of the same figure shows $\Delta$\cubi~ against
[Na/Fe], and the histogram distribution  of  [Na/Fe] is plotted in panel
$c_2$. The fact that population $a$-, $b_{1}$-, and  $b_{2}$-stars have, on
average, different sodium content, demonstrates that the oxygen  spread is
intrinsic and that population $b$ contains further sub-populations. The
position of stars belonging to the three stellar populations in the $V$
versus \cubi~ diagram is shown in panel (d).

Independent support to the fact that the star formation history of NGC\,104
is more  complex than what inferred from previous works comes from the
analysis of HB stars. Both theoretical arguments and observations have shown
that, at least in some clusters, the HB morphology is strictly related to
the presence of  multiple stellar populations \citep[e.g.\,][]{dantona08,
salaris08, marino11, gratton12}. In case of NGC\,104, there are two groups
of CN-rich and CN-poor red-HB stars  \citep{norris82}, that define two
sequences along the red HB, when observed  in appropriate filters (Milone et
al.\ 2012).

Figure~\ref{fig:HBAGB47tuc} shows data for NGC\,104. We plot in panel (a) a
CMD based on wide colour baseline, with the sample of HB stars shown as
black dots. In panels (b1) and (b2) we show the $V$ versus \cubi~ CMD and
Hess diagram, respectively. The corresponding histogram of the colour 
distribution is shown in Panel (b3) and confirms that the HB of NGC\,104 is 
not consistent with a simple stellar population. There are at least two
groups  of stars in close analogy with what observed along the RGB. A
clustered blue HB, and a more dispersed red one with the presence of at
least two components. Interestingly, stars in the red component are also
brighter than the stars in the blue portion of the HB. Moreover, a
similar conclusion can  be reached also for another cluster with
predominantly red HB morphology, namely NGC\,6366 (see
Fig.~\ref{fig:cubi}), that displays a bimodal HB made  with similar
morphology.

Though we cannot exclude that mass loss play a subtle role in shaping the
complex HB morphology, it is difficult to understand why this effect would
manifest in certain colours rather than in all CMDs. Nevertheless, we conclude 
that these results on NGC\,104 provide strong direct  evidence that the 
red-HB morphology of this GC is strictly related to  hosted multiple 
generations, and suggest that the multiple sequences discovered in the 
CMDs of other GCs studied in this paper may be related to the morphology of their HBs.

Figure~\ref{fig:HBAGB47tuc}a also reveals a large number of AGB stars,  are
marked with black asterisks. To investigate the presence of multiple stellar
populations along the AGB of 47\,Tuc, panel (c1) shows the $V$ versus \cubi ~
diagram zoomed around the AGB, with a fiducial line through the redder
component. We verticalize the AGB stars in the usual way in panel (c2),
while panel (c3) shows the histogram of the $\Delta$\cubi~ distribution. The
bimodal distribution demonstrates that also the AGB of 47\,Tuc hosts at
least two stellar populations. We emphasize that this is the first detection
of multiple sequences along a cluster AGB. Noteworthy, NGC\,104 is the only
cluster where multiple sequences have been found from the MS, to the RGB,
HB, and  finally along the AGB.

It has been suggested by \citet{norris81} that there should exist a tight
correlation between the incidence of AGB stars belonging to helium-enhanced
stellar generations and the HB  morphology. In particular, the observed
paucity (if any) in GCs of CN-strong  AGB stars could be related to the fact
that these objects would be the  progeny of He-enhanced HB stars -- those
stars populating the extended blue  tail of the HB in some GCs -- and given
these stars would evolve preferentially  as AGB-manqu{\'e} objects as a
consequence of their extremely thin envelopes,  they would miss  the AGB
stage. In this context, the ability of \cubi~  to trace the presence of
different sub-populations also along the AGB appears  useful in order to
investigate this issue in a large sample of clusters.

\section{Conclusions}\label{sec:conclusions}

This is the first of a series of papers that will present the results of the
SUMO project - a SUrvey of Multiple pOpulations in globular clusters. We
focused here on  a general description of the project, and the first general
results.  We are building a large data base of homogeneous photometry of a
large number of GCs, with the main aim of characterizing and interpreting
their multiple populations.  This will be basis for future spectroscopic
follow-up. To date, we have collected $U$-band images for a sample of
$\sim$30 clusters complementing the archive maintained by P.\,B.\,Stetson.
In this paper, we have summarized the  observational campaigns realized so
far, the data reduction strategy, and the techniques to minimize
differential reddening and polluting sources like foreground stars and
background galaxies.

We have introduced a new index,\cubi ~, defined as the difference $(U-B) -
(B-I)$, that turns out to be very efficient in disentangling  multiple
sequences along the RGB in the range of ages and metallicities covered
by selected objects. We presented results for 23 clusters, showing that
showing that:  {\it i)} they all display clear evidence of broadened or
multiple RGBs; {\it ii)} the slope of the RGB in the V-\cubi~ diagrams is very
sensitive to the mean cluster metallicity. 

Although in those clusters where multimodal RGBs are evident, small
intermediate populations always appear. These could be outliers due to
photometric errors, but we cannot exclude populations with intrinsic
intermediate properties. This continuous but inhomogeneous distribution
seems to disfavor formation mechanisms that predict bursty formation of
subpopulations.

The properties of the \cubi~ index were exploited to revise the behaviour of 
stellar populations in NGC\,104, showing an overall more complex picture
than previously discovered using HST data.

The effectiveness of the \cubi~ index was validated relating the
photometric and spectroscopic properties of RGB stars in seventeen
clusters. In particular, Na-rich and O-poor stars populate a RGB
with larger (less negative) \cubi~ index than the one defined by 
the Na-poor, O-rich sub-population. These results demonstrate 
that the our photometric approach is an alternative to the spectroscopic 
one in the identification of GC stellar populations.
In addition to this, the study of all the RGB stars provided from the 
wide field of view of the ground-based photometry, will allow us to 
investigate fundamental aspects in our understanding of multiple stellar populations, 
e.g. the radial distribution of different stellar groups and the 
real fractions of stars belonging to different stellar components. 
 Finally, we stress that other color combinations can be used to distinguish 
multiple populations in the CMDs, like \cubv ~= $(U-B) - (B-V)$.
However, the separation between the different sequences is smaller 
than that provided by \cubi~ index.

\section*{acknowledgments}

\small
Support for this work was provided by the IAC (grant P/310394) and the Education and
Science Ministry of Spain (grants AYA2010-16717). 
The research of MA is supported by Laureate Fellowship from the
Australian Research Council (grant FL110100012).
APM acknowledges the financial support from the Australian Research 
Council through Discovery Project grant DP120100475.
SC acknowledges financial support from PRIN INAF 2012 (PI: E. Carretta) and 
PRIN MIUR 2010-2011, project ``The Chemical and Dynamical Evolution of the 
Milky Way and Local Group Galaxies'', prot. 2010LY5N2T 
\normalsize

\bibliography{monelli_bibtex}

\appendix
\section{The \cubv~ as an alternative index to identify multiple RGBs}\label{sec:appendix}

\begin{centering}
\begin{figure*} 
\includegraphics[width=16cm]{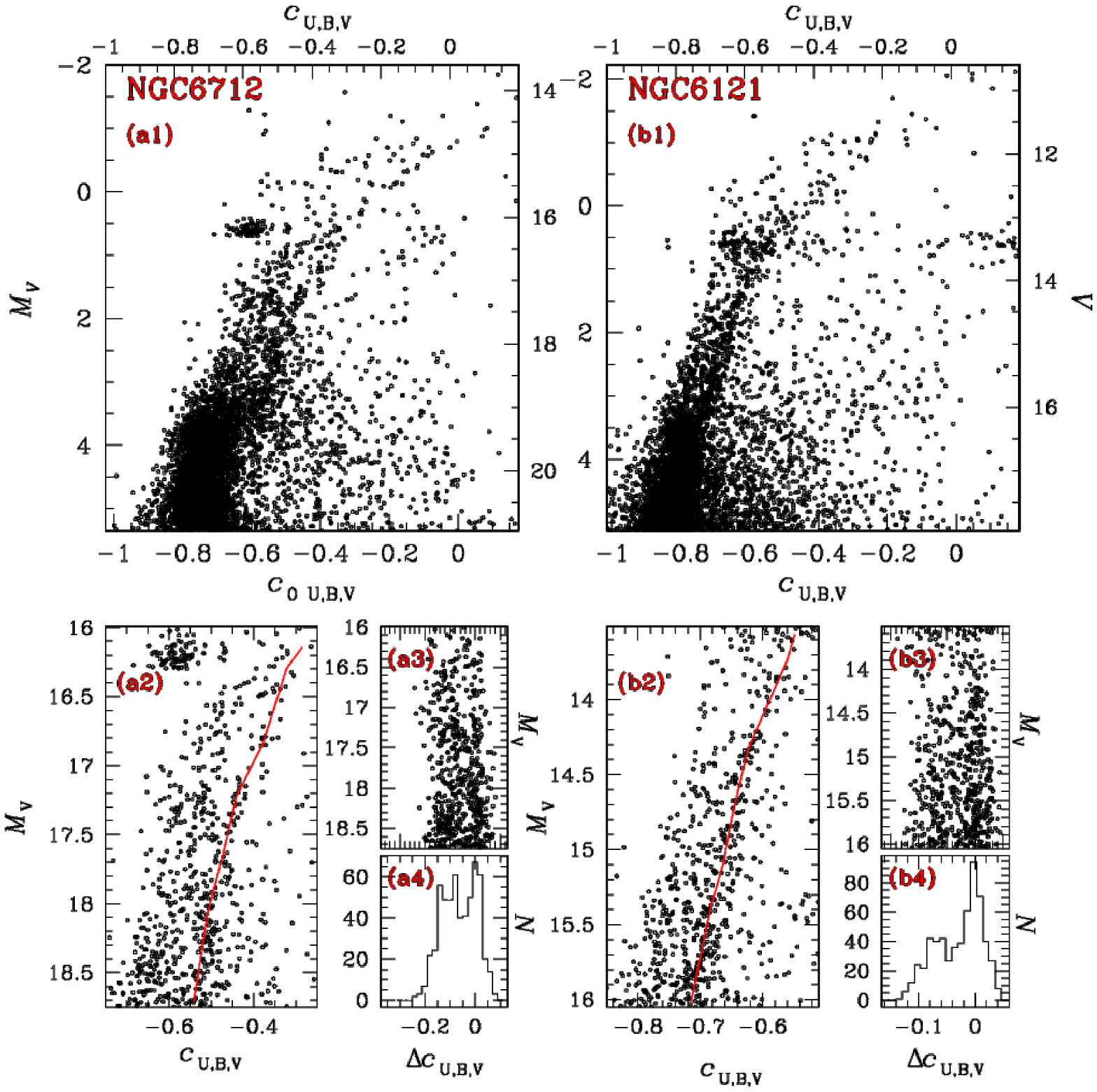}
\caption{$M_V$ {\em vs} \cubv~ of the cluster NGC\,6712, for which $I$ photometry 
is not yet available, in comparison with that of NGC\,6121.  The lower panels
show the rectified RGBs, and the $\Delta$\cubv~ colour distribution, which clearly
show hints of bimodality.}
\label{fig:cubv} 
\end{figure*}
\end{centering}

 Fig.~\ref{fig:cubv} shows the case of NGC\,6712 (panels {\em a1-a4}).
Since the $I$-band data are not yet available for this cluster, we used 
the \cubv ~= $(U-B) - (B-V)$ index. Note that the sensitivity of this index 
is smaller than that of the \cubi, as demonstrated by Fig.~\ref{fig:works}.
In fact, the colour difference between difference sequences is smaller in 
($B-V$) than in ($B-I$). Nevertheless, it allows to identify well-defined 
substructures with at least two RGBs. To support this conclusion, we compare
the \cubv~ index of NGC\,6121 (panels {\em b1-b4}), which presents clear
hints on multiple RGBs from the \cubi~ index. In particular, panels {\em a2}
and {\em b2} present a zoom in the RGB region with superimposed the ridge 
line of the reddest sequence, in analogy to what presented in Fig. 
\ref{fig:histo_proc}. The rectified RGBs are shown in {\em a3)} and {\em b3)},
and finally the histograms showing the distribution of $\Delta$cubv~ colour
differences are plotted in {\em a4)} and {\em b4)}. The latter clearly
shows bimodal distribution, thus supporting that also NGC\,6712 hosts
multiple populations.

%%%%%%%%%%%%%%%%%%%%%%%%%%%%%%%%%%%%% FIG 20 %%%%%%%%%%%%%%%%%%%%%%%%%%%%%%%%%%%
%%%%%%%%%%%%%%%%%%%%%%%%%%%%%%%%%%%%%%%%%%%%%%%%%%%%%%%%%%%%%%%%%%%%%%%%%%%%%%%%

\end{document}